\documentclass[aps,prc,twocolumn,superscriptaddress]{revtex4-1}

\usepackage{cleveref}
\usepackage{graphicx}
\usepackage{multirow}
\usepackage{amstext}
\usepackage{amssymb}
\usepackage{bbding}
\usepackage{lineno}

\begin{document}

\title{Sensitivity of stellar electron-capture rates to parent neutron number: A case study on a continuous chain of twenty Vanadium isotopes}

\author{G. W. Hitt}
\email[corresponding author ]{george.hitt@kustar.ac.ae; hitt.george@gmail.com}
\affiliation{Department of Applied Mathematics \& Sciences, Khalifa University of Science, Technology and Research, Abu Dhabi 127788, UAE}
\affiliation{Department of Nuclear Engineering, Khalifa University of Science, Technology and Research, Abu Dhabi 127788, UAE}

\author{S. Gupta}
\affiliation{Indian Institute of Technology Ropar, Nangal Road, Rupnagar (Ropar), Punjab 140 001, India}

\author{R. G. T. Zegers}
\affiliation{National Superconducting Cyclotron Laboratory, Michigan State University, East Lansing, Michigan 48824, USA}
\affiliation{Department of Physics and Astronomy, Michigan State University, East Lansing, Michigan 48824, USA}
\affiliation{Joint Institute for Nuclear Astrophysics, Michigan State University, East Lansing, Michigan 48824, USA}

\author{R. Titus}
\affiliation{National Superconducting Cyclotron Laboratory, Michigan State University, East Lansing, Michigan 48824, USA}
\affiliation{Department of Physics and Astronomy, Michigan State University, East Lansing, Michigan 48824, USA}
\affiliation{Joint Institute for Nuclear Astrophysics, Michigan State University, East Lansing, Michigan 48824, USA}

\author{C. Sullivan}
\affiliation{National Superconducting Cyclotron Laboratory, Michigan State University, East Lansing, Michigan 48824, USA}
\affiliation{Department of Physics and Astronomy, Michigan State University, East Lansing, Michigan 48824, USA}
\affiliation{Joint Institute for Nuclear Astrophysics, Michigan State University, East Lansing, Michigan 48824, USA}

\author{B. A. Brown}
\affiliation{National Superconducting Cyclotron Laboratory, Michigan State University, East Lansing, Michigan 48824, USA}
\affiliation{Department of Physics and Astronomy, Michigan State University, East Lansing, Michigan 48824, USA}
\affiliation{Joint Institute for Nuclear Astrophysics, Michigan State University, East Lansing, Michigan 48824, USA}

\author{A. L. Cole}
\affiliation{Physics Department, Kalamazoo College, Kalamazoo, Michigan 49006, USA}

\author{S. Shams}
\affiliation{Department of Applied Mathematics \& Sciences, Khalifa University of Science, Technology and Research, Abu Dhabi 127788, UAE}

\date{\today}

\begin{abstract}
Gamow-Teller (GT) strength distributions (B(GT)) in electron-capture (EC) daughters stemming from the parent ground state are computed with the shell-model in the full $pf$-shell space, with quasi-particle random-phase approximation (QRPA) in the formalism of Krumlinde and M\"oller and with an Approximate Method (AM) for assigning an effective B(GT). These are compared to data available from decay and charge-exchange (CE) experiments across titanium isotopes in the $pf$-shell from $A=43$ to $A=62$, the largest set available for any chain of isotopes in the $pf$-shell. The present study is the first to examine B(GT) and the associated EC rates across a particular chain of isotopes with the purpose of examining rate sensitivities as neutron number increases. EC rates are also computed for a wide variety of stellar electron densities and temperatures providing concise estimates of the relative size of rate sensitivities for particular astrophysical scenarios. This work underscores the astrophysical motivation for CE experiments in inverse kinematics for nuclei away from stability at the luminosities of future Radioactive Ion Beam Facilities.
\end{abstract}


\maketitle

\section{Introduction\label{intro}}

The $pf$-shell region of the nuclear chart is an important frontier for experimental and theoretical nuclear astrophysics investigations. Stellar fusion cycles, in both massive stars and on the surfaces of compact objects, synthesize light elements into heavier ones and terminate with the production of $^{56}$Ni near the center of the $pf$-shell. A wide variety of explosive scenarios, such as type-I x-ray bursts\cite{Schatz2006601,Parikh2013225}, type IA supernovae\cite{0067-0049-125-2-439,Seitenzahl21022013} and core-collapse supernovae\cite{PhysRevLett.91.201102,Langanke2014305}, neutron star mergers (see Ref.\cite{2041-8205-738-2-L32}, also Ref.\cite{arXiv.1509.07628} and references therein) and the quiescent processes regulating the thermal balance of neutron star crusts prior to x-ray superbursts\cite{0004-637X-662-2-1188,PhysRevLett.101.231101,refId0,10.1038.nature.12757} are candidates for various nucleosynthesis processes that push isotopic abundances out of the $pf$-shell region and into the proton- and neutron-rich \textit{terra incognita} of the nuclear chart. 

In these scenarios, which usually take place under nearly electron-degenerate conditions, the evolutionary path and astronomical observables (e.g. the x-ray emission determined by the thermal balance of the accreting neutron star crust) are sensitive to the details of the nuclear weak interaction. As a consequence, the nuclear structure pertinent to $\beta$-decay and electron-capture (EC) processes of neutron-rich nuclei is a primary ingredient necessary to model these processes accurately.

In this paper, we report on a comparative study examining the distribution of Gamow-Teller (GT) strength (B(GT)) in EC daughters, stemming from the parent ground state, to those available from decay and charge-exchange (CE) experiments across Vanadium (V) isotopes in the $pf$-shell from $A=43$ to $A=62$. This corresponds to astrophysical electron fractions ($Y_{e})$ of $\sim0.53$ to $\sim0.37$. Accordingly, $Z/A$ is used to elucidate all our results. The EC daughter is Titanium ($^{A}$Ti) for a $^{A}$V parent so in each case, the theoretical B(GT) is determined in the daughter Titanium (Ti) nucleus with large-scale shell-model calculations using two different effective interactions, with quasi-particle random-phase approximation (QRPA) in the formalism of Krumlinde and M\"oller, and with an approximate parametrization assigning an effective GT state in the daughter spectrum. Last, the associated absolute EC rates and their sensitivities are estimated and compared for these isotopes, for a wide variety of astrophysical electron density ($\rho Y_{e}$) and temperature conditions.

\begin{table*}
\caption{\label{Table1}Initial and final states, EC Q-values, available data and extent of SM calculations for B(GT:$^{A}$Ti) determination}
\begin{ruledtabular}
\begin{tabular}{clccccccccc}
$i$ & $g.s.$ & $f$ & $f$-g.s. & $GT$-states & Q$_{\textrm{EC}}$\footnote{Q-values based on AME2012\cite{ame2012}} & $\beta$ & (d,$^{2}$He) & $E_{x}^{max}$(GXPF1a)(MeV) & $E_{x}^{max}$(KB3G)(MeV)\\
\hline
$^{43}$V & (7/2$^{-}$)\footnote{cases where ENSDF makes no assignment, then AME2012\cite{ame2012} recommendation is used} & $^{43}$Ti & 7/2$^{-}$  &  (5/2$^{-}$,7/2$^{-}$,9/2$^{-}$)  & +11.41(4) &   &   & 10.9 & 10.5 \\
$^{44}$V & (2)$^{+}$   & $^{44}$Ti & 0$^{+}$                   & (1,2,3)$^{+}$     & +13.43(18)    & \cite{Hagberg1997183} &   & 12.7 & 12.5 \\ 
$^{45}$V & 7/2$^{-}$   & $^{45}$Ti & 7/2$^{-}$   & 5/2$^{-}$,7/2$^{-}$,9/2$^{-}$   &  +7.129(8)    & \cite{HORNSHOJ19824} &   & 7.6 & 7.5 \\ 
$^{46}$V & 0$^{+}$     & $^{46}$Ti & 0$^{+}$                         & 1$^{+}$     &  +7.05239(9)  & \cite{PhysRevLett.73.396} &   & 12.6 & 12.4 \\ 
$^{47}$V & 3/2$^{-}$   & $^{47}$Ti & 5/2$^{-}$   & 1/2$^{-}$,3/2$^{-}$,5/2$^{-}$   &  +2.93060(15) & \cite{Fifield1973516} &   & 8.1 & 8.2 \\ 
$^{48}$V & 4$^{+}$     & $^{48}$Ti & 0$^{+}$         & 3$^{+}$,4$^{+}$,5$^{+}$     &  +4.0150(10)  & \cite{Z.Phys.A273.405} &   & 7.7 & 7.8 \\ 
$^{49}$V & 7/2$^{-}$   & $^{49}$Ti & 7/2$^{-}$   & 5/2$^{-}$,7/2$^{-}$,9/2$^{-}$   &  +0.6019(8)   & \cite{Z.Phys.251.375} &   & 6.2 & 6.5 \\ 
$^{50}$V & 6$^{+}$     & $^{50}$Ti & 0$^{+}$         & 5$^{+}$,6$^{+}$,7$^{+}$     &  +2.2068(9)   &   & \cite{PhysRevC.68.031303} & 12.2 & 12.6 \\ 
$^{51}$V & 7/2$^{-}$   & $^{51}$Ti & 3/2$^{-}$   & 5/2$^{-}$,7/2$^{-}$,9/2$^{-}$   &  -2.4718(10)  &   & \cite{PhysRevC.71.024603} &  9.5 & 9.8 \\ 
$^{52}$V & 3$^{+}$     & $^{52}$Ti & 0$^{+}$         & 2$^{+}$,3$^{+}$,4$^{+}$     &  -1.975(7)    &   &   & 10.0 & 10.2 \\ 
$^{53}$V & 7/2$^{-}$   & $^{53}$Ti & (3/2)$^{-}$   & 5/2$^{-}$,7/2$^{-}$,9/2$^{-}$ &  -5.02(10)    &   &   & 8.8 & 8.7 \\ 
$^{54}$V & 3$^{+}$     & $^{54}$Ti & 0$^{+}$         & 2$^{+}$,3$^{+}$,4$^{+}$     &  -4.30(13)    &   &   & 9.9 & 9.4 \\ 
$^{55}$V & (7/2$^{-}$)$^{\textrm{b}}$ & $^{55}$Ti & (1/2)$^{-}$ & (5/2$^{-}$,7/2$^{-}$,9/2$^{-}$) &  -7.48(16) &   &   & 8.9 & 7.8 \\ 
$^{56}$V & 1$^{+}$     & $^{56}$Ti & 0$^{+}$         & 0$^{+}$,1$^{+}$,2$^{+}$     &  -6.92(20)    &   &   & 13.0 & 12.4 \\ 
$^{57}$V & (7/2$^{-}$) & $^{57}$Ti & (5/2$^{-}$) & (5/2$^{-}$,7/2$^{-}$,9/2$^{-}$) & -10.4(3)      &   &   & 8.4 & 7.2 \\ 
$^{58}$V & (1$^{+}$)   & $^{58}$Ti & 0$^{+}$       & (0$^{+}$,1$^{+}$,2$^{+}$)     &  -9.2(4) &   &   & 10.1 & 8.9 \\ 
$^{59}$V & (5/2$^{-}$$^{\textrm{b}}$, & $^{59}$Ti & (5/2$^{-}$)$^{\textrm{b}}$ & (3/2$^{-}$,5/2$^{-}$,7/2$^{-}$) & -12.2(4) &   &   &  &  \\
         & 3/2$^{-}$   &           &  & (1/2$^{-}$,3/2$^{-}$,5/2$^{-}$)            &          &   &   & &\\
		 & 7/2$^{-}$$^{\footnote{shell-model predictions for the ground state in a $pf$ model space disagree with ENSDF and AME2012}}$)  &           &  & (5/2$^{-}$,7/2$^{-}$,9/2$^{-}$)            &          &   &   & 5.9 & 5.0 \\
$^{60}$V & (3$^{+}$)$^{\textrm{b}}$   & $^{60}$Ti & 0$^{+}$       & (2$^{+}$,3$^{+}$,4$^{+}$)     & -10.9(5) &   &   & 7.5 & 7.6 \\ 
$^{61}$V & (3/2$^{-}$$^{\textrm{b}}$, & $^{61}$Ti & (1/2)$^{-}$ & (1/2$^{-}$,3/2$^{-}$,5/2$^{-}$) & -14.2(11) &   &   &  &  \\ 
         & 5/2$^{-}$)  &           &  & (3/2$^{-}$,5/2$^{-}$,7/2$^{-}$)            &          &   &   & &\\
		 & 7/2$^{-}$$^{\textrm{c}}$)  &           &  & (5/2$^{-}$,7/2$^{-}$,9/2$^{-}$)            &          &   &   & 15.4 & 18.5 \\
$^{62}$V & (3$^{+}$)$^{\textrm{b}}$   & $^{62}$Ti & 0$^{+}$       & (2$^{+}$,3$^{+}$,4$^{+}$)     & -12.9(8) &   &   &  17.9 & 21.5 \\ 

\end{tabular}
\end{ruledtabular}
\end{table*}

Though individual nuclei have been the subject of EC rate sensitivity studies recently\cite{PhysRevC.86.015809,PhysRevLett.112.252501,PhysRevC.92.024312}, the distribution of B(GT) and EC rates across a chain of isotopes allows structural effects on rate sensitivities to be revealed. This work is thus complementary to recent EC rate sensitivity studies such as Refs.\cite{PhysRevC.86.015809,0004-637X-816-1-44}. 

Table \ref{Table1} summarizes the isotopes and data considered. EC Q-values are calculated using AME2012 \cite{ame2012}. The V$\rightarrow$Ti-chain is primarily chosen for a case study because the B(GT) distributions available from decay or CE data are the largest for any chain of isotopes in the $pf$-shell, making it useful for extracting systematic trends with increasing neutron number. It is noteworthy that $^{44}$Ti is included in the set of isotopes studied here since, among other nuclei, $^{44}$Ti production in core-collapse occurs exclusively during alpha-rich freeze-out which is sensitive to variations in pre-collapse models \cite{1991ApJ...368L..31W}. Furthermore, as it is a long-lived gamma-ray emitter, is observable in the remnant and so its abundance allows for more direct connection between models and observation (e.g. Ref. \cite{nature11473} and references therein). Though all data available are necessarily for proton-rich cases (decay) or stable cases (CE), the relatively large amount of data for adjacent Ti daughter isotopes and trends in their associated EC rates allows the clear highlighting of:

\begin{enumerate}

\item the general limitations imposed by Q-values on using decay data to constrain model-based determinations of B(GT) and EC rates,
\item the relative scarcity of CE data, which overcome Q-value restrictions, to better constrain models,
\item the increase in EC rate sensitivity, with increasing neutron number, occurring at $Q = -m_{e}c^{2}$,
\item the introduction of new and/or Pauli blocking of single-particle configurations in the GT transition and
\item the pressing need for further CE experiments in inverse kinematics toward greater neutron-richness to reduce uncertainties in the nuclear inputs to astrophysical simulations.

\end{enumerate}

In this context, we stress that $Y_{e}$ evolution is crucial to both nucleosynthetic flows and resultant yields, as well as hydrodynamical evolution via the pressure of degenerate electrons, and accordingly our sensitivity estimation plots utilize $Z/A$, to elucidate changing nuclear structure and electroweak rate effects with increasing neutronization in astrophysical sites. As we show, even 1-3 additional neutrons from stability will have a major impact on the EC rate sensitivity, especially for core-collapse supernovae conditions, and should provide strong motivation for further CE experiments.

\section{Methodology\label{method}}

The methodology of the present work is very similar to that presented in Ref.\cite{PhysRevC.86.015809} which we briefly review here. 

\subsection{Structure and reaction rate calculation}

The B(GT) distributions from shell-model (SM) calculations are determined using the NuShellX code \cite{nushellx} in the full $pf$-shell space, using both GXPF1a \cite{gxpf1a} and KB3G \cite{Poves2001157} effective interactions. The strength is then scaled by applying a well-known quenching factor $(0.744)^{2}$\cite{PhysRevC.53.R2602} that accounts for strength lost to excitations outside the model space. SM calculations are extended in daughter excitation energy ($E_{x}$) until the summed B(GT) strength expected from an independent particle model estimation is exhausted. The $E_{x}$ of the highest GT state calculated in the daughter sufficient to achieve this is listed down right-most columns in Table \ref{Table1}. The B(GT) distributions from QRPA are calculated as per Ref.\cite{Krumlinde1984419}, in the manner used for global determinations \cite{MOLLER19901,MOLLER1997131}, up to $E_{x} \sim 25-30$ MeV. EC rates associated with each B(GT) distribution are calculated using the code developed in Ref.\cite{0004-637X-662-2-1188} using the FFN formalism \cite{1980ApJS...42..447F,1982ApJ...252..715F,1982ApJS...48..279F,1985ApJ...293....1F} where the capture rate $\lambda$ is given by

\begin{equation}
\label{rateequation}
\lambda = \ln (2)\sum_j \frac{f_{j}(T,\rho,U_{F})}{(ft)_{j}}
\end{equation}

\noindent in terms of the $ft$-value and phase space integral $f_{j}$ for the $j^{th}$ GT state in the daughter spectrum. The $f_{j}$ integral is solved by the code \cite{0004-637X-662-2-1188} for Fermi-Dirac-distributed, continuum electrons in the stellar plasma, as a function of electron density $\rho Y_{e}$ and degeneracy $U_{F}/k_{B}T$, where at $T=0$, $U_{F} = 0.511\lbrace[1.018(\rho_{6}Y_{e})^{2/3}+1]-1 \rbrace$ (MeV). $U_{F}(T)$ is the electron chemical potential ($\mu_{e}(T)$) which at $T=0$ is just the Fermi energy ($\epsilon_{F}$). At low temperature, $U(T)$ is essentially a step function allowing only those GT daughter states satisfying $E_{x}-Q_{\textrm{EC}} \leq \mu_{e}(T)=U(T)$ to contribute to the EC rate. For purposes of selecting data sets, discussed below, we use the rough approximation $\epsilon_{\rm{F}}+2kT$ derived from this inequality as a $Q$-value threshold criterion (hereafter, TC). The threshold effect is most pronounced at lower densities and temperatures where a relative few, low-lying states account for the accesible strength and small changes in the B(GT) distribution lead to large differences in EC rate. At higher temperatures, the Fermi surface smears out to about $\sim kT$ and differences between B(GT) distributions can become less important. At higher densities, the EC rate is largely determined by the summed B(GT) and the detailed distribution of B(GT) no longer has a major influence on EC rate differences. In addition to EC rates determined from the above nuclear structure calculations, we also compare with EC rates determined using an Approximate method (``AM'' hereafter). As introduced in Ref. \cite{1985ApJ...293....1F} and studied in detail for core-collapse scenarios in Refs. \cite{PhysRevLett.90.241102,0004-637X-816-1-44}, the AM is a parametrization which assigns a single, effective excitation energy and GT transition strength to a given EC daughter nucleus.

Fermi(F)- and GT-type operator strengths from the nuclear structure theories determine the $ft$-value for the $j^{th}$ state through the relation

\begin{equation}
\label{ftequation}
(ft)_{j} = \frac{K/g_{v}^{2}}{B(F_{j})+(g_{A}/g_{V})^{2}B(GT_{j})}
\end{equation}

\noindent where constants $K/g_{v}^{2} = 6143 \pm 2$s \cite{PhysRevC.79.055502} and $(g_{A}/g_{V})^{2} = -1.2694 \pm 0.0028$ \cite{0954-3899-37-7A-075021}. Likewise, $ft$- and $B(F,GT)$-values from decay or CE data are interchanged with the same equation. Since $B(F)$ is mostly model-independent and relatively easy to determine, it is not considered in the determination of theoretical $ft$-values or EC rates in this work. The focus of the present work is to examine various methods for determining the $B(GT)$ contribution to the rate, so the $B(F)$ term in Eq.(\ref{ftequation}) is only considered when removing a $B(F)$ component from decay data. Thus, all rates considered are determined from $B(GT)$ distributions only. Absolute rates determined in this way from Eq.(\ref{rateequation}) are not necessarily representative of the real stellar environment, but make for a convenient and completely self-consistent way to estimate rate sensitivities from rate ratios, whether between two rates derived from different nuclear structure theories or from a theory versus a data set.

Whenever available, the ENSDF \cite{ensdf} spin assignments are used for EC parent and daughter nucleus ground states. For cases where ENSDF makes no spin assignment, the Atomic Mass Evaluation 2012 (AME2012) \cite{ame2012} recommendation is used. In three cases, for captures on $^{55,59,60}$V, differences between ENSDF or AME2012 and shell-model calculations produce significant changes to the derived EC rates. In the case of $^{55}$Ti, GXPF1a agrees with the g.s. spin assignment from ENSDF (1/2$^{-}$), placing the first GT state (a 5/2$^{-}$ state) at 0.899 MeV in $^{55}$Ti, but KB3G determines 5/2$^{-}$ for the g.s. and therefore assigns it GT strength. In the case of $^{59}$V, ENSDF recommends two possible g.s. assignments (5/2$^{-}$,3/2$^{-}$) but reports possible shape-coexistence, while AME2012 recommends 5/2$^{-}$, based on systematics. Shell-model calculations with GXPF1a and KB3G predict 7/2$^{-}$ for the ground state. Similarly, for $^{61}$V, ENSDF recommends two possible g.s. assignments (3/2$^{-}$,5/2$^{-}$) but reports possible shape-coexistence, while AME2012 recommends 3/2$^{-}$. Again, shell-model calculations with GXPF1a and KB3G predict 7/2$^{-}$ for the ground state. In both cases, the shell-model predictions for the ground states were used for estimating overall rate sensitivities to ensure self-consistency.

Decay data for $^{44}$V $\beta ^{+}$ decay and $^{45}$V $\beta ^{+}$ decay are taken from Refs.\cite{Hagberg1997183,HORNSHOJ19824}, respectively. Decay data for $^{46}$V $\beta ^{+}$ decay, $^{47}$V $\beta ^{+}$ decay, $^{48}$V EC decay and $^{49}$V EC decay are taken from Refs.\cite{PhysRevLett.73.396,Fifield1973516,Z.Phys.A273.405,Z.Phys.251.375}, respectively. CE determinations of B(GT) in the two stable V-parents are taken from the $^{50}$V($d$,$^{2}$He) \cite{PhysRevC.68.031303} and $^{51}$V($d$,$^{2}$He) \cite{PhysRevC.71.024603} reactions.



Estimating the EC rate sensitivity across a chain of isotopes at a given ($\rho Y_{e}$, $T_{9}$) is performed as follows; 
\begin{enumerate}
  \item A thermodynamic grid point ($\rho Y_{e}$, $T_{9}$) and isotopic chain (parent $Z$-value) of interest is chosen and an associated TC value ($\epsilon_{F}(\rho Y_{e}) + 2kT$) is calculated.
  \item The EC rate based on any $\beta$-decay data sets where $Q_{\rm{EC}} < \textrm{TC}$ is excluded from further analysis, since GT strength accessible in the daughter is unknown, compromising the validity of comparison. Unhindered by this, CE data is always included.
  \item Analysis categories (see subsection below for details), bounded by an interval of neutron number values, are determined based on reaction $Q$-value, the onset of intruder states and Pauli blocking
  \item For isotopes where their associated data remains following application of the TC (the ``sample set''), as in (2.) above, the log-ratio $\log(\lambda_{t,d} / \lambda_{KB3G})$ between each theoretical rate ($\lambda_{t}$) or data-based rate ($\lambda_{d}$) is computed against that of the SM with KB3G ($\lambda_{KB3G}$) for each structural category, for cross comparison.
  \item EC rate sensitivities are determined by computing average values of the log-ratio of rates, with averages spanning isotopes within each analysis category. Differences in average log-ratios are compared across categories for significance.
\end{enumerate}

\begin{figure}
\includegraphics[bb = 30bp 0bp 430bp 450bp,clip,scale=0.49]{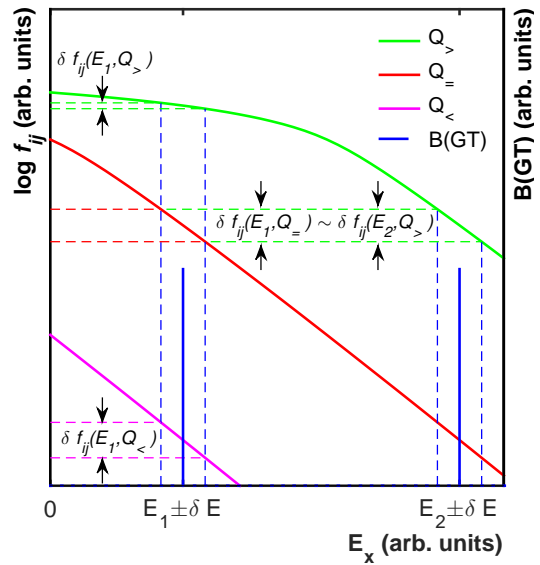}
\caption{\label{qsample} A schematic showing how uncertainty $\delta E$ in the distribution of B(GT) (right axes) over daughter $E_{x}$ translates into uncertainty $\delta f_{ij}$ in the phase space (left axes) available for the reaction, for three representative reaction $Q$-values: $Q_{>}$, where $Q > -m_{e}c^{2}$, $Q_{=}$, where $Q = -m_{e}c^{2}$ and $Q_{<}$, where $Q < -m_{e}c^{2}$.}
\end{figure}

\subsection{Definition of analysis categories}

We define three analysis categories based on structural considerations, as prescribed in step 3 of the above procedure, which are described as follows;
\begin{enumerate}
  \item Parent $N < N_{A}$ (labeled ``$<N_{A}$'' hereafter), $N_{A}$ being determined by the $N$-value where with increasing $N$, the $Q$-value falls below the reaction threshold $-m_{e}c^{2}$. For the V$\rightarrow$Ti chain, this occurs at $N = 28$ ($^{51}$V), so that $20 \leq N \leq 27$ ($^{43-50}$V) cases are assigned to the $<N_{A}$ category. An important feature of this category, as discussed below, is significantly decreased EC rate sensitivity to low-lying B(GT) distribution, relative to cases in neutron-rich categories.
  \item Parent $N \geq N_{A}$ but $N < N_{B}$ (labeled ``$N_{A}\rightarrow N_{B}$'' hereafter) where $N_{B}$ is determined by the $N$-value where, with increasing $N$, the $g_{9/2}$ (or states built upon it) single-particle level begins to intrude into the $pf$-shell spectrum. For the V$\rightarrow$Ti chain, this occurs at $N \sim 34$ ($^{57}$V) based on a comparison of SM and QRPA single-particle occupancies, so that $28 \leq N \leq 33$ ($^{51-56}$V) cases are assigned to the $N_{A}\rightarrow N_{B}$ category. An important feature of this category is the relative comparability of model-spaces across the various nuclear structure theories used in the B(GT) determination.
  \item Parent $N \geq N_{B}$ (labeled ``$>N_{B}$'' hereafter) which for the V$\rightarrow$Ti chain contains $^{57-62}$V) cases, up to the presumable edge of the $pf$-shell ($N = 40$) in a simple IPM picture. This category could be further subdivided at $N = 40$ where in such a picture, the neutron $f_{5/2}$ level is filled, but this leaves too few cases (3) upon which to base a meaningful estimate of the rate sensitivity. Thus, the important feature of the $>N_{B}$ category is the relative incomparability of model spaces across the structure theories used. We emphasize that while we perform SM calculations in this region to support an argument, the results themselves do not represent the real B(GT) well because of the incomplete model space and so should not be used for other purposes.
\end{enumerate}

\begin{figure*}
\centering
\includegraphics[bb = 0bp 0bp 575bp 730bp,clip,scale=0.41]{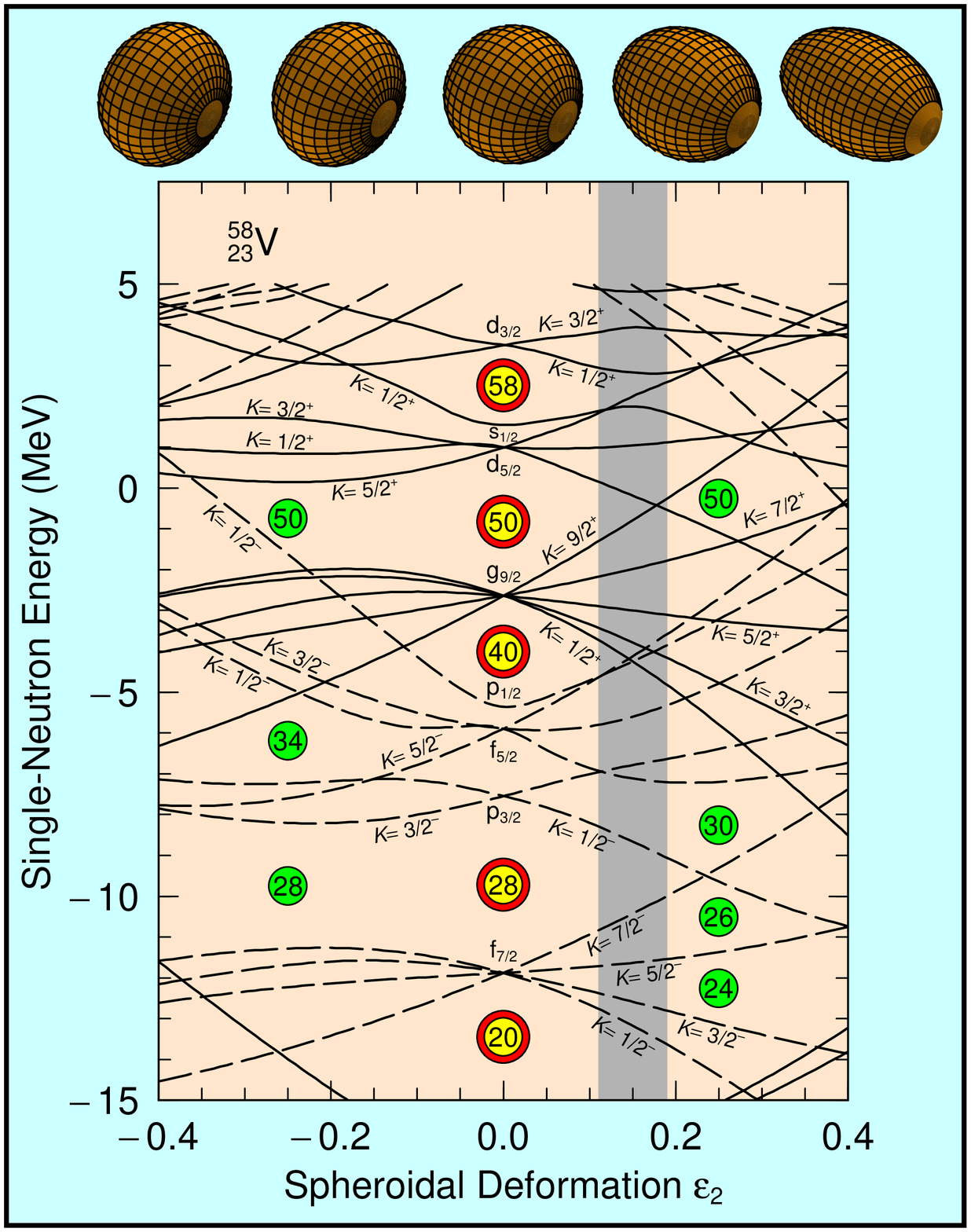}
\includegraphics[bb = 0bp 0bp 575bp 730bp,clip,scale=0.41]{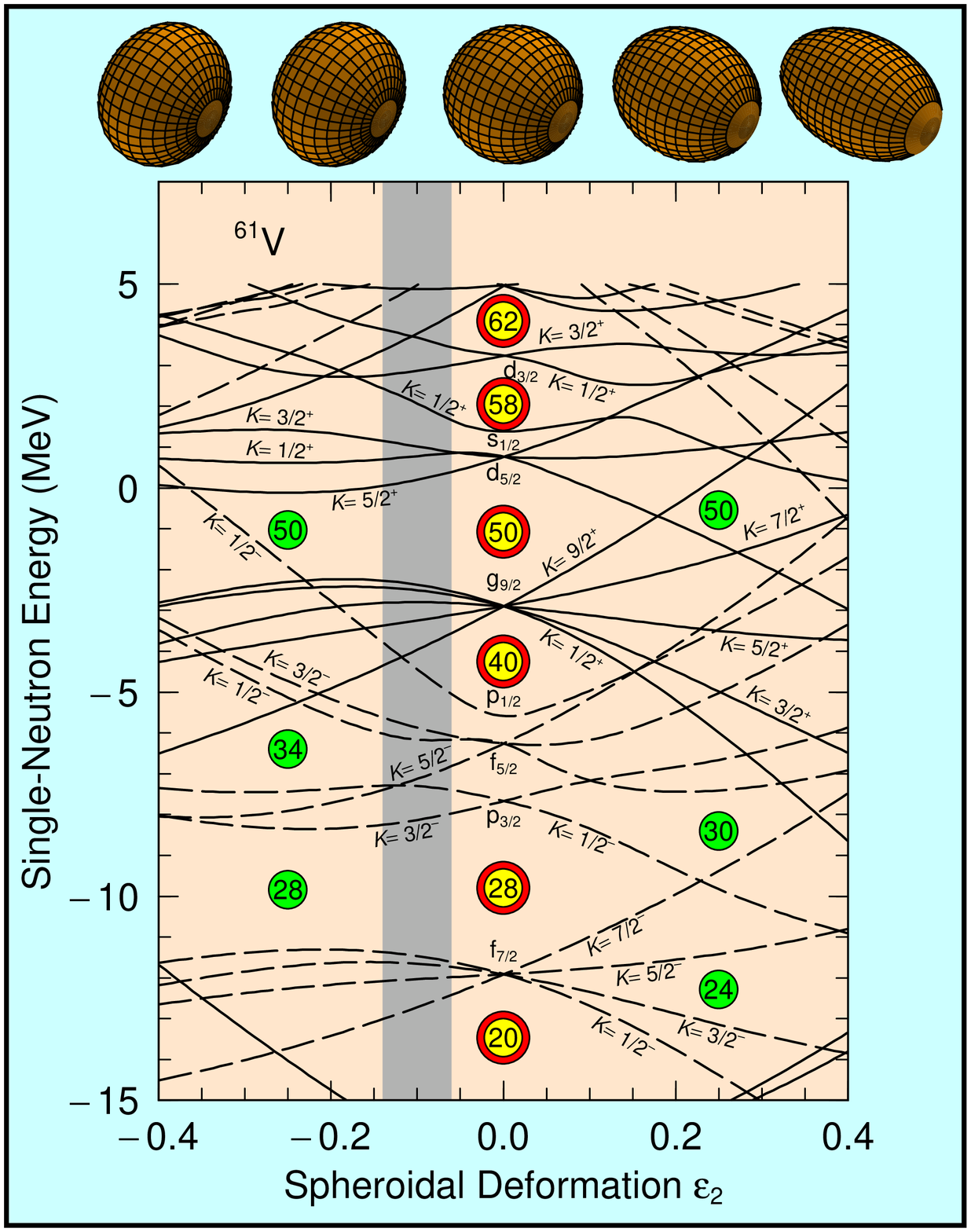}
\caption{\label{levn} Nilsson neutron diagrams for $^{58}$V (left) and $^{61}$V (right) used in the construction of their respective QRPA states. Darkened vertical bands indicate ground state deformation and highlight the energy ordering of single-neutron states. States are labeled by $K^{\pi}$, the suitable quantum number for the rotational symmetry involved, and parity. For spherical symmetry (zero deformation), these states reduce to those used in shell-model configurations, indicated along the center vertical in $l_{J}$-notation, the orbital and total intrinsic angular momentum projections.}
\end{figure*}

\begin{table*}
\caption{\label{qrpa} Neutron single quasi-particle levels, quantum numbers and occupation factors for $^{58-61}$V cases in QRPA calculations.}
\parbox{.22\linewidth}{
\centering
\begin{minipage}{\linewidth}
\begin{center}
$^{58}$V
\end{center}
\begin{ruledtabular}
\begin{tabular}{ccc}
level \#& $K^{\pi}$&$V^{2}$\\
\hline
18 & 3/2$^{-}$ & 0.5858 \\
19 & 1/2$^{-}$ & 0.2109 \\
20 & 5/2$^{-}$ & 0.1981 \\
21 & 1/2$^{+}$ & 0.1277 \\
22 & 3/2$^{+}$ & 0.1041 \\

\end{tabular}
\end{ruledtabular}
\end{minipage}
}
\hfill
\parbox{.22\linewidth}{
\centering
\begin{minipage}{\linewidth}
\begin{center}
$^{59}$V
\end{center}
\begin{ruledtabular}
\begin{tabular}{ccc}
level \#& $K^{\pi}$&$V^{2}$\\
\hline
18 & 3/2$^{-}$ & 0.5547 \\
19 & 1/2$^{-}$ & 0.1930 \\
20 & 5/2$^{-}$ & 0.1800 \\
21 & 1/2$^{+}$ & 0.1780 \\
22 & 3/2$^{+}$ & 0.1309 \\

\end{tabular}
\end{ruledtabular}
\end{minipage}
}
\hfill
\parbox{.22\linewidth}{
\centering
\begin{minipage}{\linewidth}
\begin{center}
$^{60}$V
\end{center}
\begin{ruledtabular}
\begin{tabular}{ccc}
level \#& $K^{\pi}$&$V^{2}$\\
\hline
18 & 1/2$^{-}$ & 0.7822 \\
19 & 3/2$^{-}$ & 0.7196 \\
20 & 9/2$^{+}$ & 0.2218 \\
21 & 1/2$^{-}$ & 0.1848 \\
22 & 7/2$^{+}$ & 0.1256 \\

\end{tabular}
\end{ruledtabular}
\end{minipage}
}
\hfill
\parbox{.22\linewidth}{
\centering
\begin{minipage}{\linewidth}
\begin{center}
$^{61}$V
\end{center}
\begin{ruledtabular}
\begin{tabular}{ccc}
level \#& $K^{\pi}$&$V^{2}$\\
\hline
18 & 1/2$^{-}$ & 0.7787 \\
19 & 3/2$^{-}$ & 0.7483 \\
20 & 1/2$^{-}$ & 0.2575 \\
21 & 9/2$^{+}$ & 0.1677 \\
22 & 7/2$^{+}$ & 0.1109 \\

\end{tabular}
\end{ruledtabular}
\end{minipage}
}
\end{table*}

Fig.\ref{qsample} shows the justification for creating the $<N_{A}$ category. From Eq.\ref{rateequation} and Eq.\ref{ftequation}, the contribution to the EC rate by a given GT-state is determined by the product $B(\textrm{GT}_{j}) \times f_{j}$ (right and left axes, respectively), so that if there is an uncertainty $\delta E$ in the location of a given GT-state in the daughter excitation spectrum $E_{x}$, there is a corresponding, large uncertainty in the phase space $\delta f_{ij}$ available to the reaction for that channel. The phase space is plotted in Fig.\ref{qsample} as function of $E_{x}$ in the daughter for $Q$-values of $Q > -m_{e}c^{2}$ ($Q_{>}$, green), $Q = -m_{e}c^{2}$ ($Q_{=}$, red) and $Q < -m_{e}c^{2}$ ($Q_{<}$, magenta). There are multiple demonstrations of the importance of low-lying part of the B(GT) for the overall EC rate and rate sensitivity in the literature (e.g. \cite{PhysRevC.80.014313,PhysRevC.92.024312,PhysRevLett.112.252501}) which is due to the much larger amplitude of the phase space function at low $E_{x}$, a feature that cases of all possible $Q$-values have in common. However, the \textit{degree} of rate uncertainty caused by $\delta E$ is not the same for all three $Q$-value cases.

The degree of rate uncertainty caused by $\delta E$ stems from the two contributions to the phase space: a relatively slowing decreasing part on $0 < E_{x} \lesssim Q$ coming from terms describing the density of electron states in the continuum and a more rapidly decreasing part on $E_{x} \gtrsim Q$ coming from the tail of the electron Fermi-Dirac distribution. The second part is present for all three $Q$-value cases, but the first part is only present for cases where $Q > -m_{e}c^{2}$ ($Q_{>}$, green). Thus, the EC rate uncertainty due to $\delta E$ for a low-lying GT state from decay data sets ($Q_{>}$) will not give the same relative rate uncertainty as a low-lying state from CE data (corresponding to ($Q_{=}$, red) and ($Q_{<}$, magenta) curves, for forward or inverse-kinematics) because the B(GT)s are multiplied by phase space functions with fundamentally different $E_{x}$ dependence. For this reason, one expects a sudden increase in EC rate sensitivity with respect to increasing $N$, occuring at an $N$-value approximately where $Q = -m_{e}c^{2}$, such that cases above and below the boundary value should be separately categorized and compared.


Fig.\ref{levn} and Table \ref{qrpa} show the justification for creating the $N_{A}\rightarrow N_{B}$ category. Of the methods used to determine the B(GT), SM and AM use either the full $pf$-shell model space directly or, in the case of AM rates, are parameterizations based on fits to SM calculations in the full $pf$-shell \cite{PhysRevLett.90.241102} which bases both on an assumption of spherical symmetry. QRPA calculations alternatively use states built upon deformed ground states, so that the effective model spaces and the degrees of freedom available in the B(GT) determination can be quite different from SM and AM methods. In the V$ \rightarrow$Ti chain, QRPA predictions for ground state deformations are negligibly small up to $^{56}$V. Starting with $^{57}$V however, ground state deformation and the resulting angular momentum raises the energy of states built on the spherical $p_{1/2}$ state, regardless of prolate (e.g. $^{58}$V, Fig.\ref{levn} (left)) or oblate (right) deformation. Similarly, most $K$-states built on the spherical $g_{9/2}$ have lower energies for both prolate and oblate deformations, so that the $N = 40$ magic number consistently disappears for cases above $A = 56$. Table \ref{qrpa} shows the degree to which this occurs by showing the corresponding occupation factors for single-neutron quasi-particle levels near the neutron Fermi surface. Each level listed holds 2 neutrons, so that in the case of $^{58}$V (N = 35), the last neutron should be in level \# 18 in a simple IPM picture. Instead, the occupation factor for level \# 18 is only 0.5858, with higher levels having significant occupation, including positive-parity states built on the intruder $g_{9/2}$ state outside the $pf$-shell. Thus, the model spaces for SM and AM are comparable with QRPA only up to $A = 56$ and while we perform SM calculations to demonstrate this, the results of SM for $A > 56$ should not be used for other purposes.

The onset of Pauli blocking further justifies separating the $>N_{B}$ category from $N_{A}\rightarrow N_{B}$ cases at lower $N$. In a simple IPM picture, GT transitions in the $pf$-shell should be completely blocked above $N = 40$. However, in the same picture, the $f_{5/2}$ neutron single-particle state is partially occupied starting at $N = 34$. Thus, differences in the treatment of blocking (and unblocking) will preclude the direct comparability of B(GT)s produced by different models. In the case of AM rates, there is no accounting for blocking, so that one can anticipate that AM consistently and increasingly overestimates EC rates starting at $N = 34$. SM calculations account for blocking of course, and allow for unblocking within the $pf$-shell model space through residual interactions. QRPA calculations also account for blocking and unblocking, but do so while  further including quasi-particle states built on single-particle states from outside the $pf$-shell.

\subsection{Representative subset of isotopes and thermodynamic grid points}
The B(GT) and EC rates for all 20 isotopes in the V$\rightarrow$Ti chain have been calculated, with rates determined over a large thermodynamic grid with $10 \leq \rho Y_{e} \leq 10^{14}$ g/cm$^{3}$ and $0.01 \leq T_{9} \leq 100$. In Figs.\ref{fig3} and \ref{fig4}, we present selected isotopes as a representative overview, showing (top row) the deduced B(GT) distributions (a-d) and sum strength (e) in select Ti EC daughters based on shell-model with (a) GXPF1a and (b) KB3G $pf$-shell effective interactions, (c) QRPA, and (d) beta-decay or CE data, where available. Two example cases are chosen to represent each of the identified analysis categories: $^{44,50}\textrm{V}\rightarrow^{44,50}\textrm{Ti}$ for the $<N_{A}$ group, $^{51,54}\textrm{V}\rightarrow^{51,54}\textrm{Ti}$ for the $N_{A}\rightarrow N_{B}$ group, and $^{57,62}\textrm{V}\rightarrow^{57,62}\textrm{Ti}$ for the $>N_{B}$ group. Beneath each panel (bottom row) are the corresponding absolute EC rates, including that determined using the AM, as a function of temperature, where yellow cross-hatching denotes the astrophysically relevant temperature range, for densities of (a) 10$^{7}$ g/cm$^{3}$ and (b) 10$^{9}$ g/cm$^{3}$ and (c) 10$^{11}$ g/cm$^{3}$. Fig.\ref{fig5} shows the logarithm of EC rate ratios (log-ratios) relative to that based on KB3G, both with respect to EC parent neutron number $N$ and electron fraction $Y_{e}$, for three astrophysical scenarios: (1) Type Ia supernovae (10$^{7}$ g/cm$^{3}$, 3$\times$10$^{9}$K) flame front or massive star pre-collapse phase, (2) core-collapse supernovae (10$^{9}$ g/cm$^{3}$, 10$\times$10$^{9}$K), and (3) a density at which neutrons drip out of these nuclei in the neutron star crust but less dense that that for pycnonuclear fusion (10$^{11}$ g/cm$^{3}$, 0.4$\times$10$^{9}$K). Consistent with Step 2 in the analysis, data-based values are only used in calculations and plotted in Fig.\ref{fig5} if they meet the TC: seven, four and three such cases in scenarios (1),(2) and (3), respectively.

\section{Results \& Discussion \label{resdis}}

 \begin{figure*}
 \centering
 \null\hfill%
 \quad\quad{$^{44}$V} \hfill {$^{50}$V} \hfill {$^{51}$V}\quad\quad%
 \hfill\null\par\medskip
 \includegraphics[bb = 0bp 0bp 385bp 517bp,clip,scale=0.315]{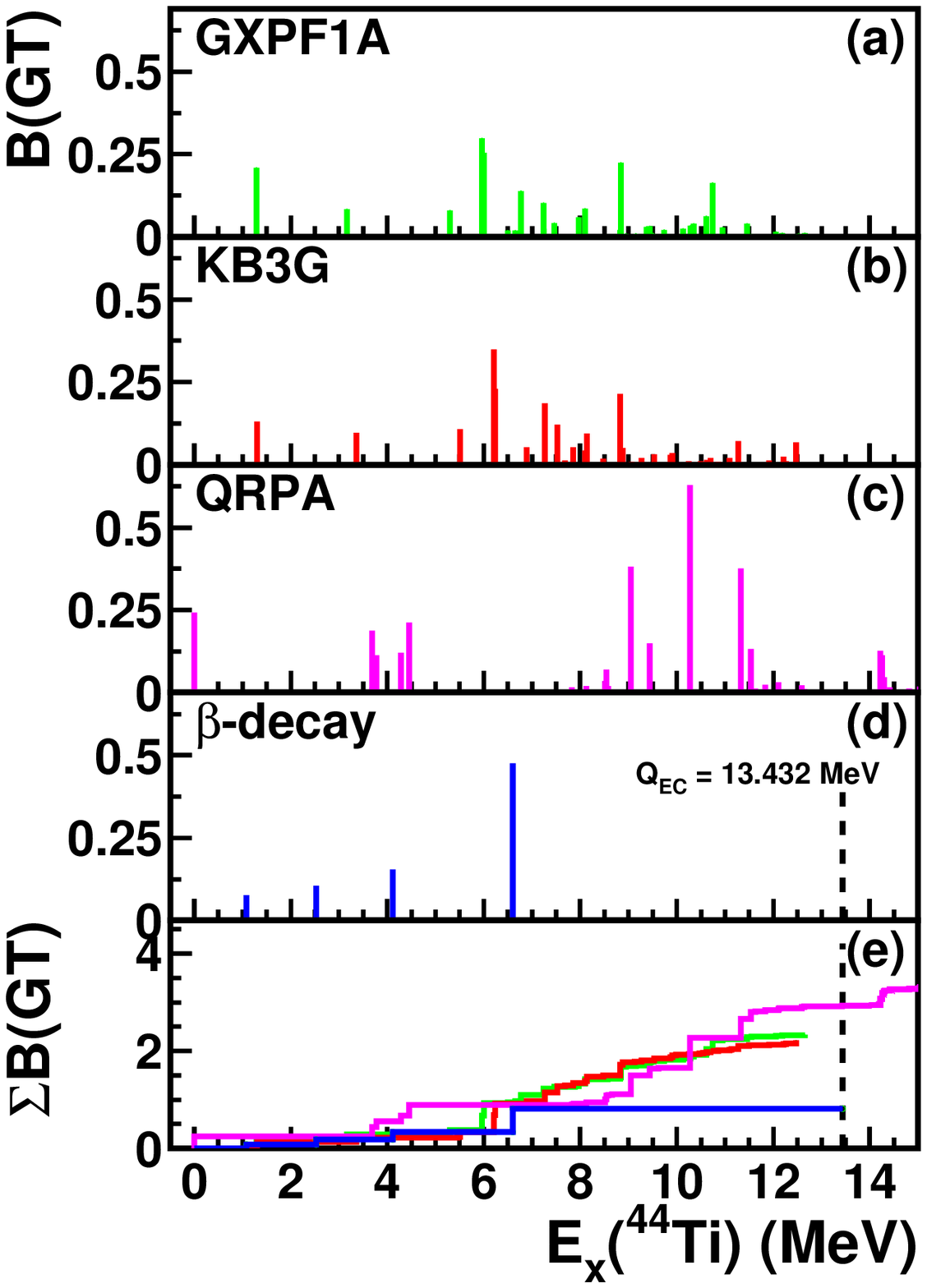}
 \includegraphics[bb = 0bp 0bp 385bp 517bp,clip,scale=0.315]{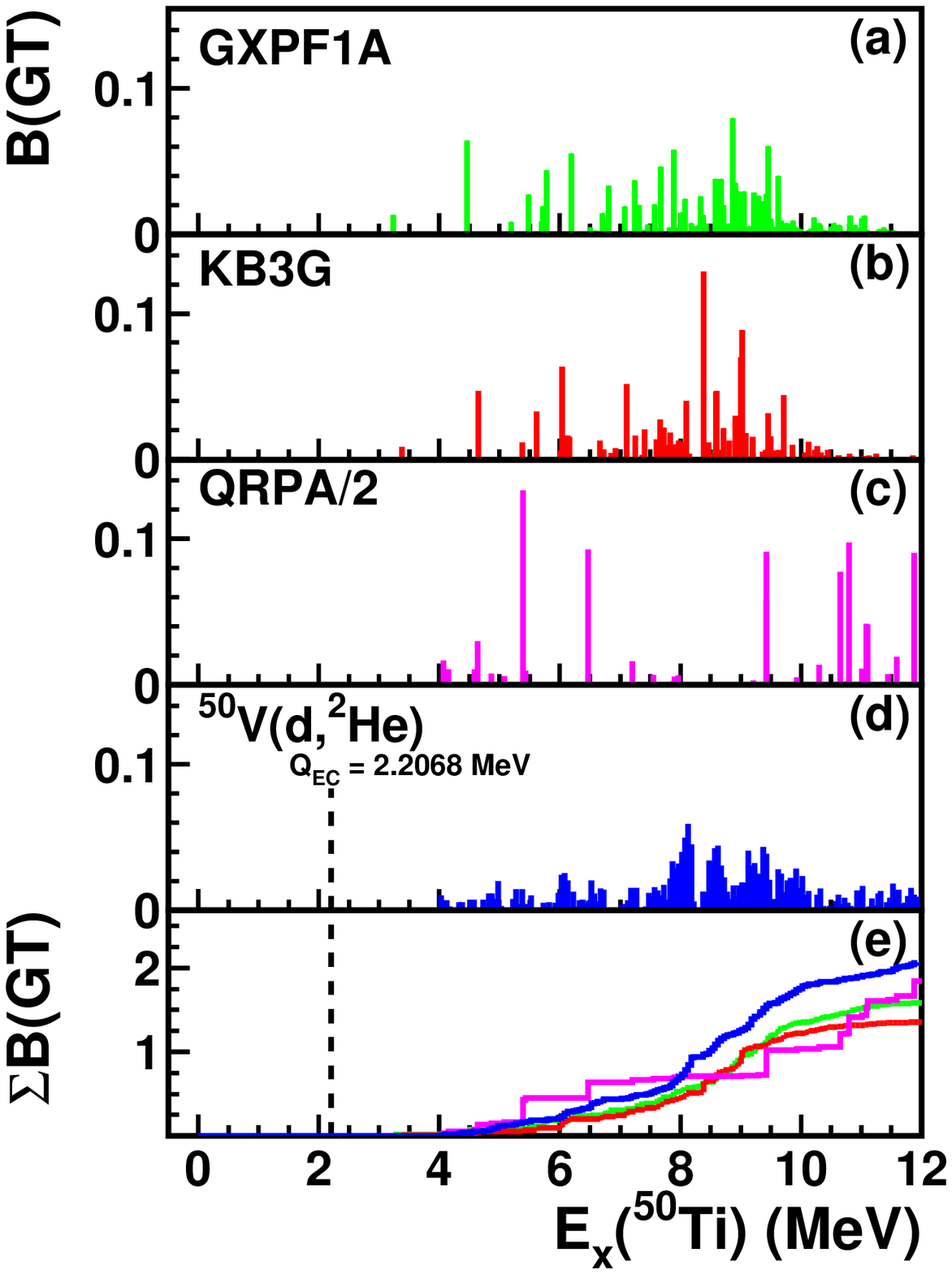}
 \includegraphics[bb = 0bp 0bp 385bp 517bp,clip,scale=0.315]{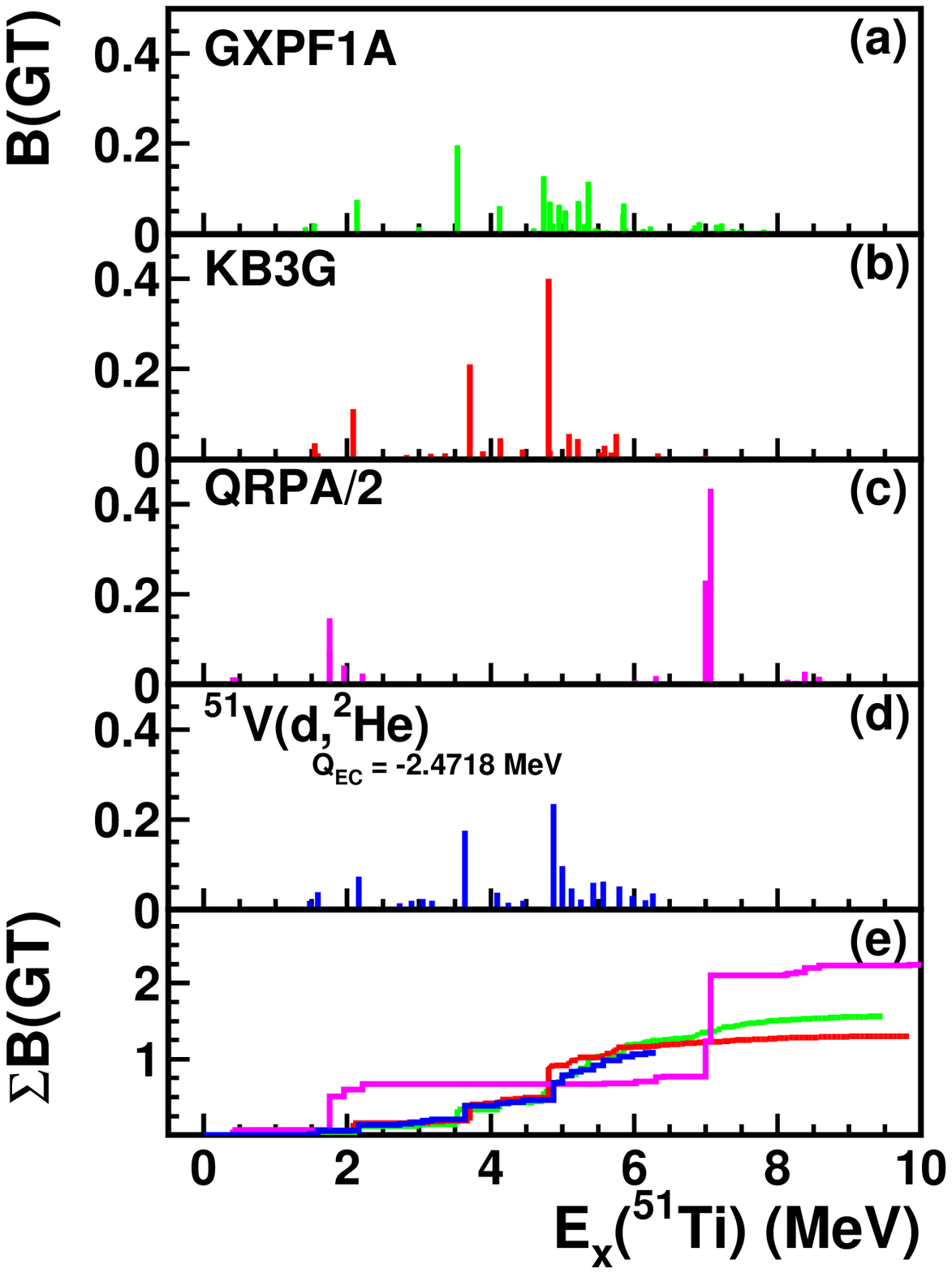}
 
 \includegraphics[bb = 0bp 0bp 385bp 667bp,clip,scale=0.315]{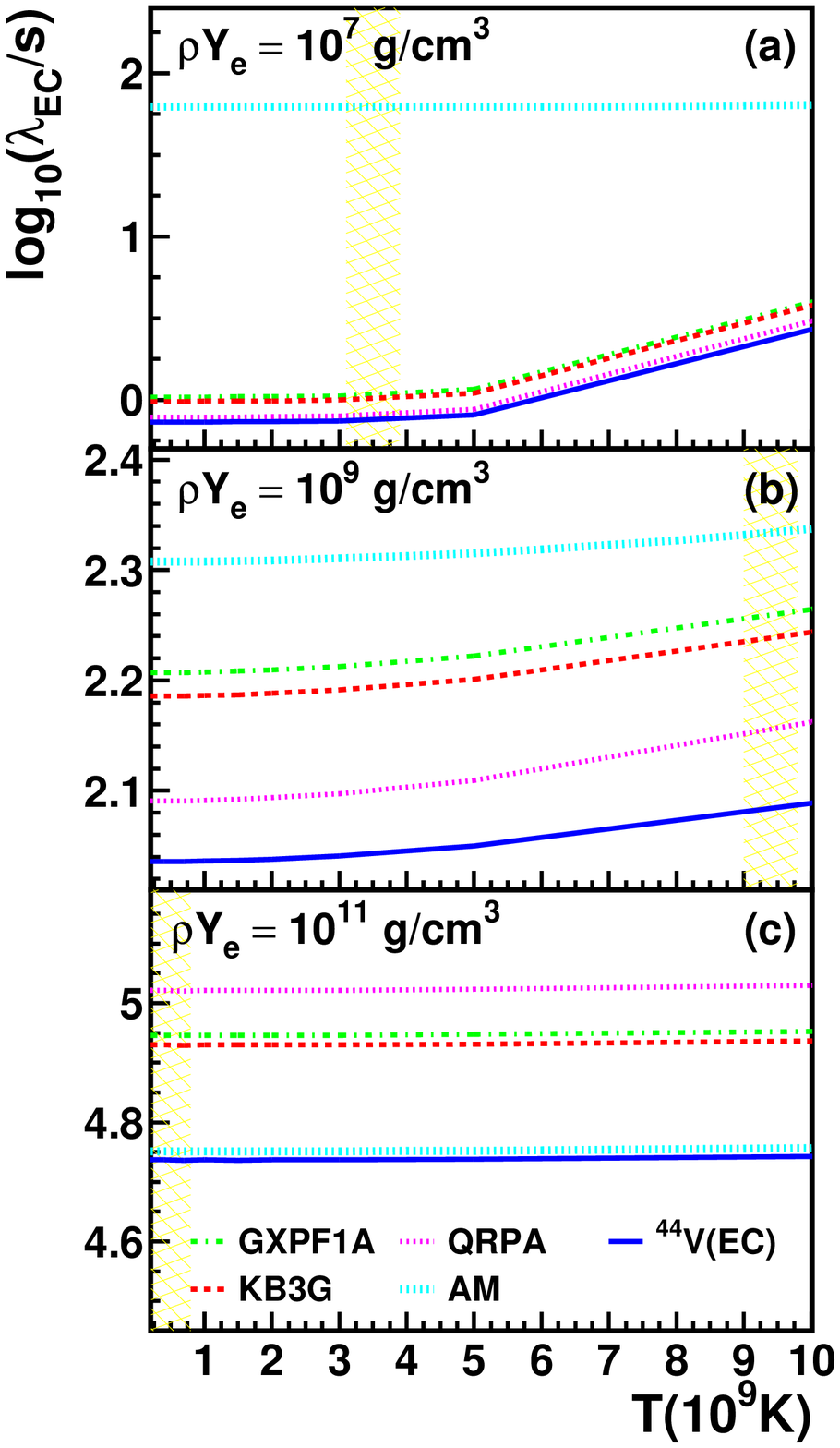} 
 \includegraphics[bb = 0bp 0bp 385bp 667bp,clip,scale=0.315]{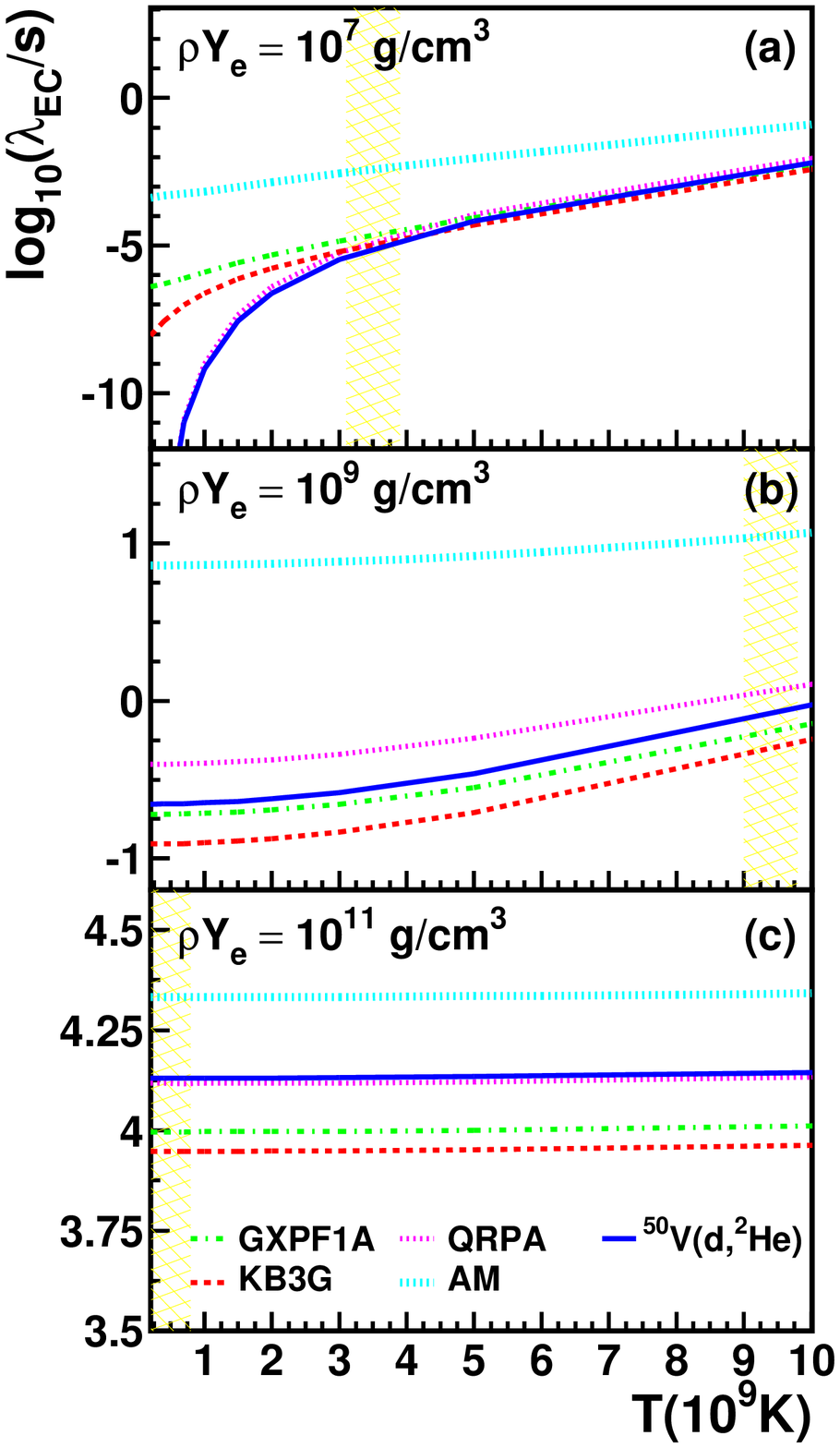}
 \includegraphics[bb = 0bp 0bp 385bp 667bp,clip,scale=0.315]{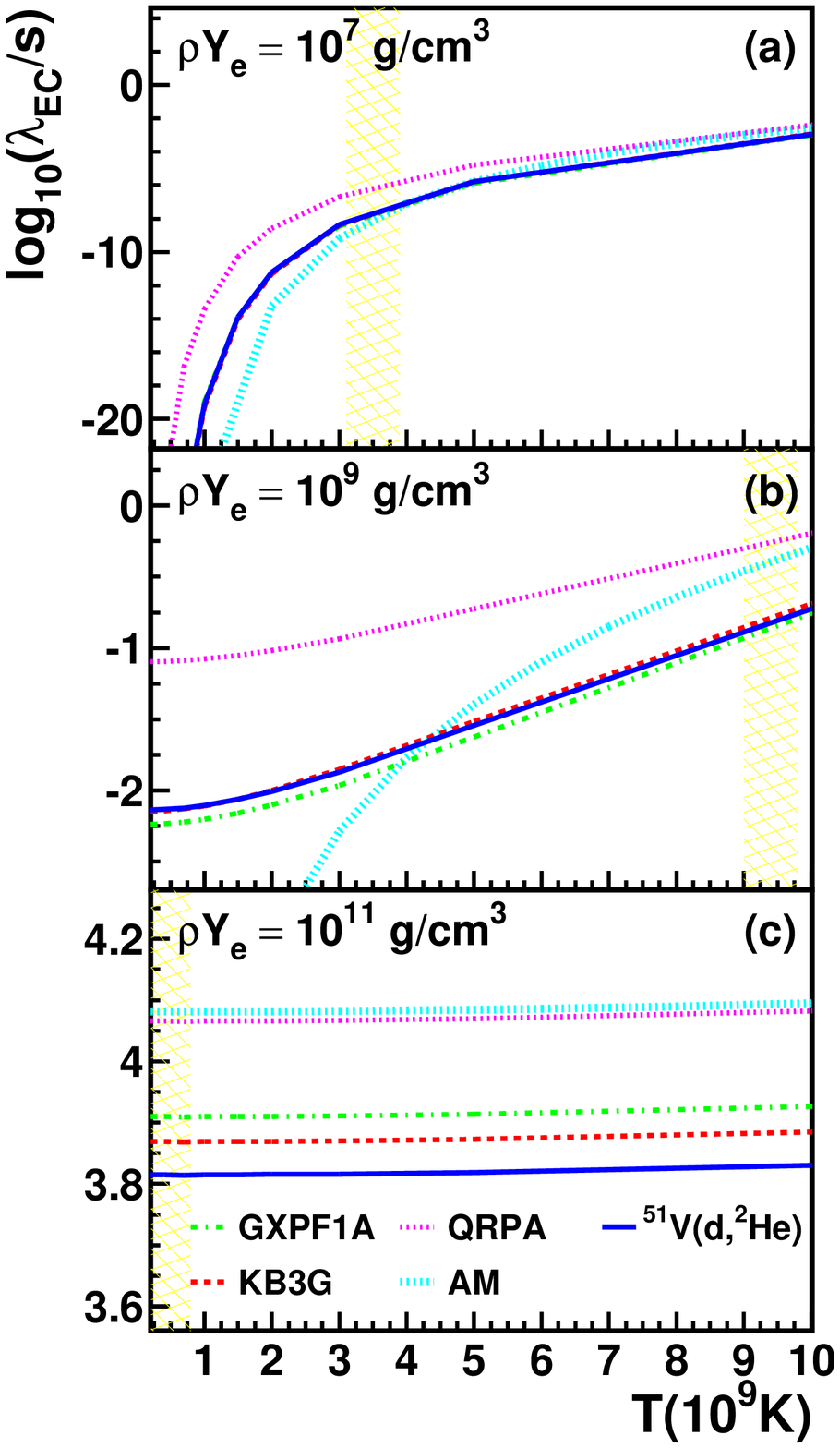}
 
 \caption{\label{fig3} The deduced B(GT) (top row) distributions (a-d) and sum strength (e) in select Ti nuclei, from ECs on the $^{43-62}$V isotopic chain, based on shell-model with (a) GXPF1a and (b) KB3G $pf$-shell effective interactions, (c) QRPA, and beta-decay or CE data (d), where available. Here we show example results from $^{44,50,51}$V$\rightarrow ^{44,50,51}$Ti cases (left, center and right, respectively). Beneath each panel (bottom row) are the corresponding EC rates, including that determined using the AM, as a function of temperature for stellar electron densities of (a) 10$^{7}$, (b) 10$^{9}$ and (c) 10$^{11}$ g/cm$^{3}$. The temperature regions used to determine select EC rate ratios for sensitivity analysis are highlighted in yellow cross-hatching and represent relevant temperatures for a variety of astrophysical scenarios.}
 \end{figure*}

 \begin{figure*}
 \centering
 \null\hfill%
 \quad\quad{$^{54}$V} \hfill {$^{57}$V} \hfill {$^{61}$V}\quad\quad%
 \hfill\null\par\medskip
 
 \includegraphics[bb = 0bp 0bp 385bp 517bp,clip,scale=0.315]{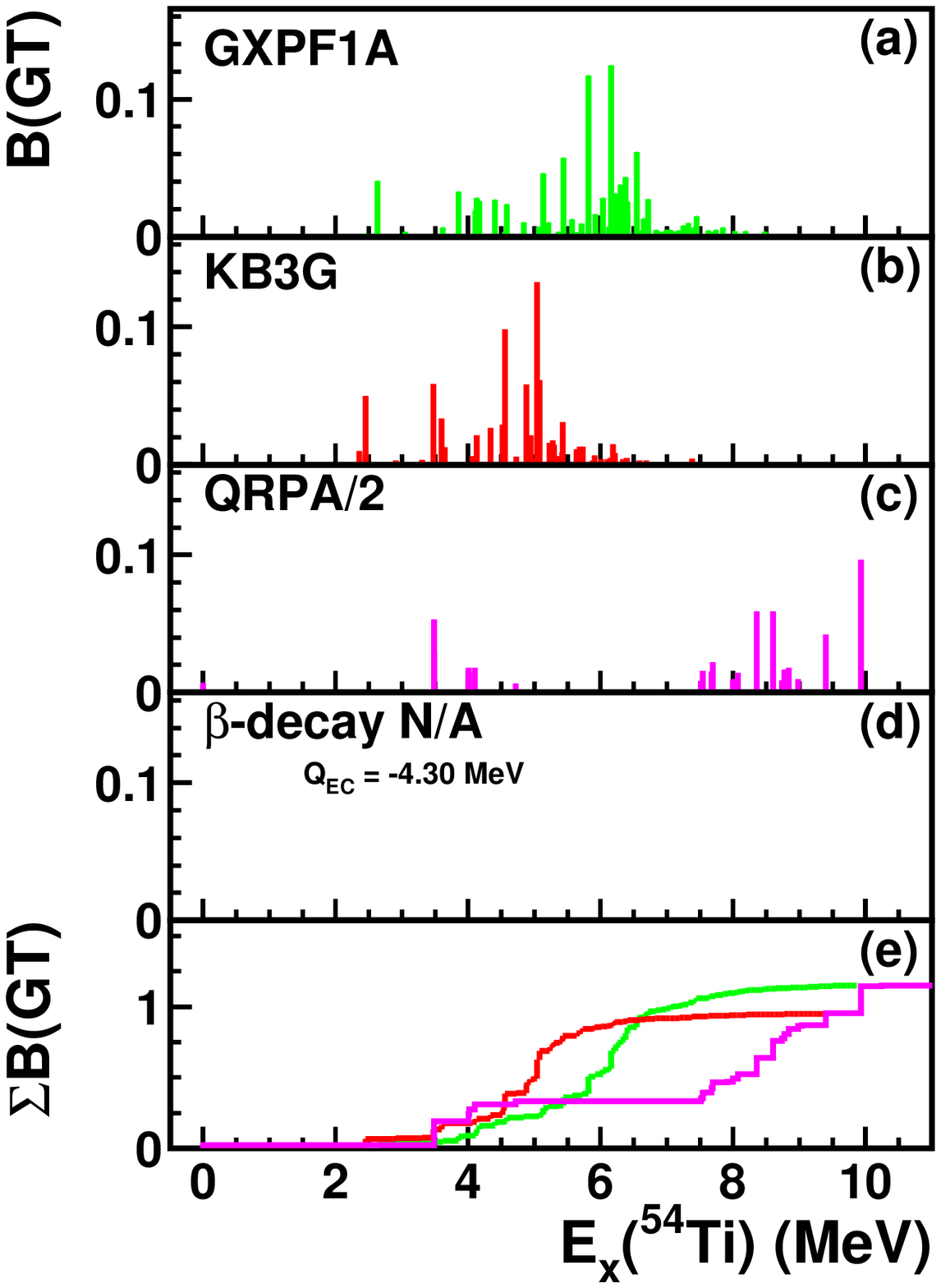}
 \includegraphics[bb = 0bp 0bp 385bp 517bp,clip,scale=0.315]{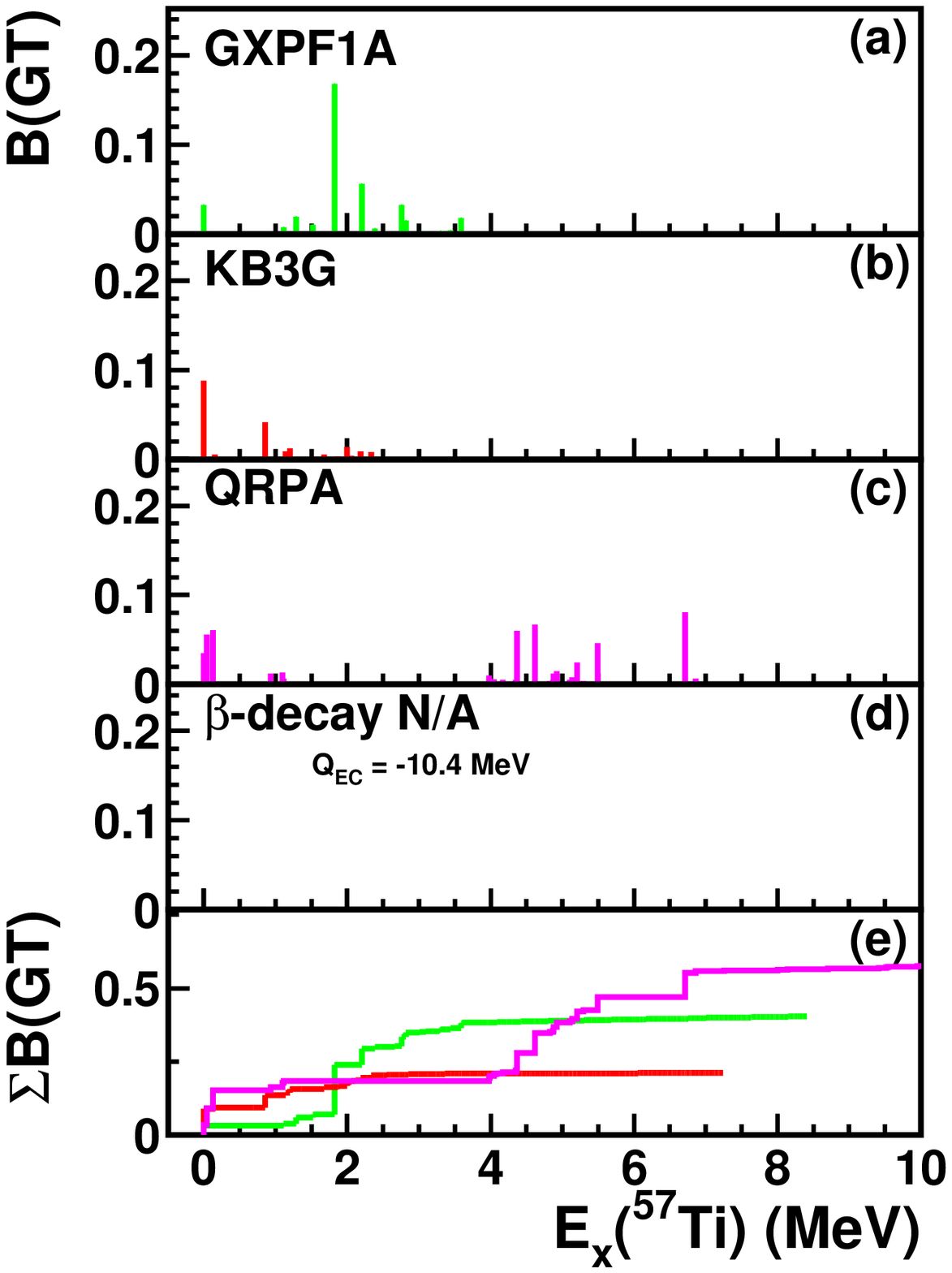}
 \includegraphics[bb = 0bp 0bp 385bp 517bp,clip,scale=0.315]{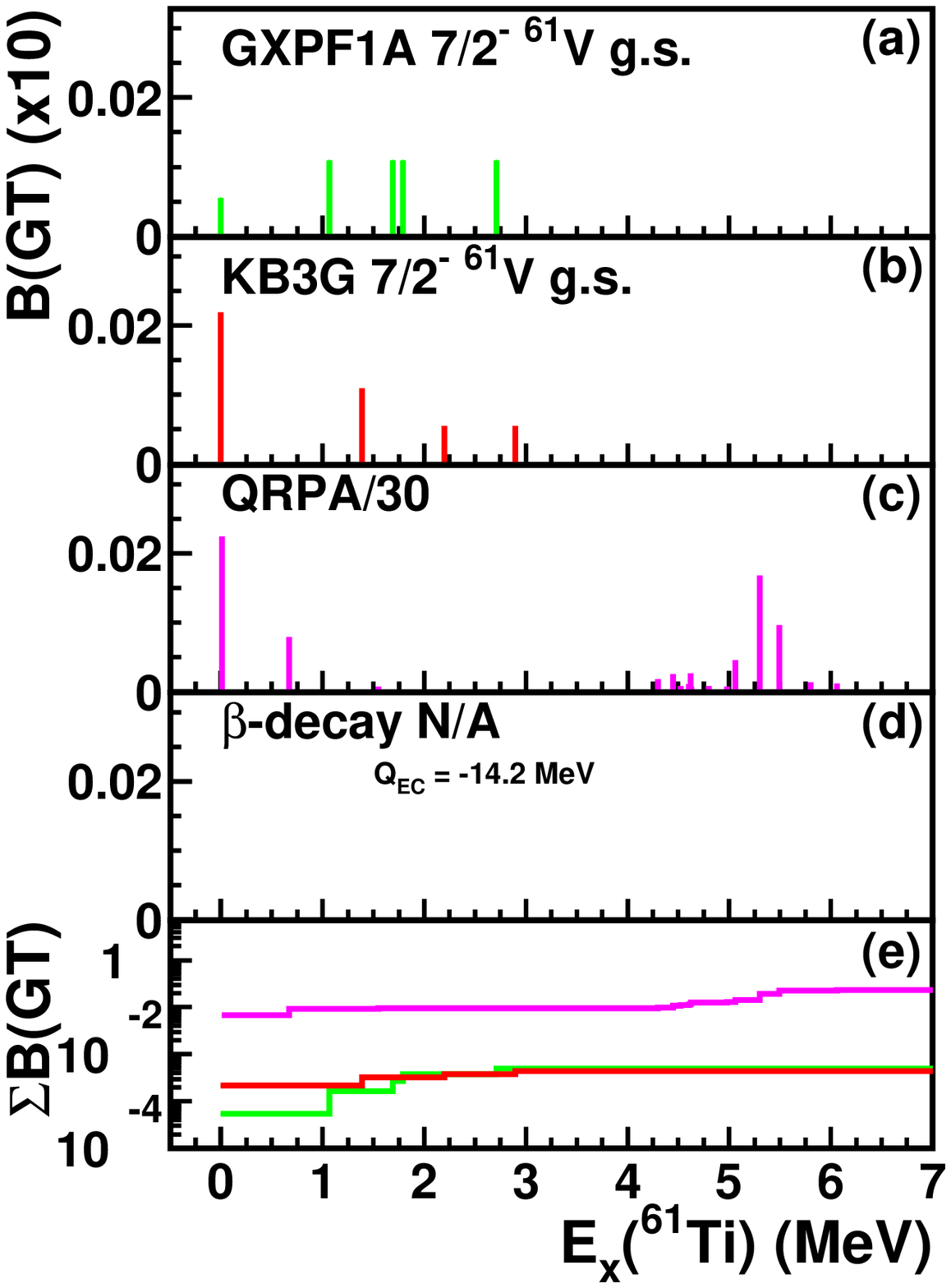}
 
 \includegraphics[bb = 0bp 0bp 385bp 667bp,clip,scale=0.315]{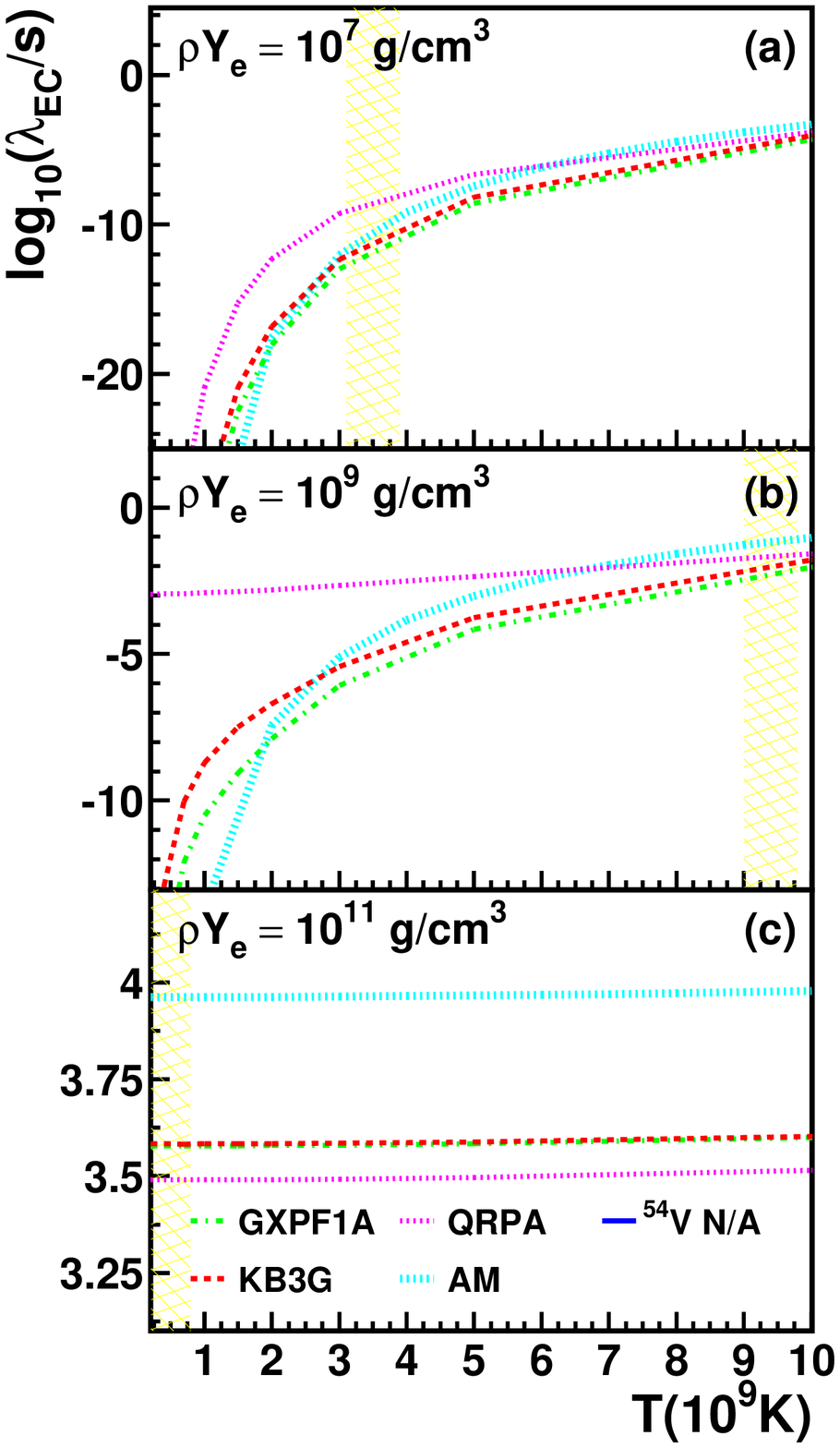}
 \includegraphics[bb = 0bp 0bp 385bp 667bp,clip,scale=0.315]{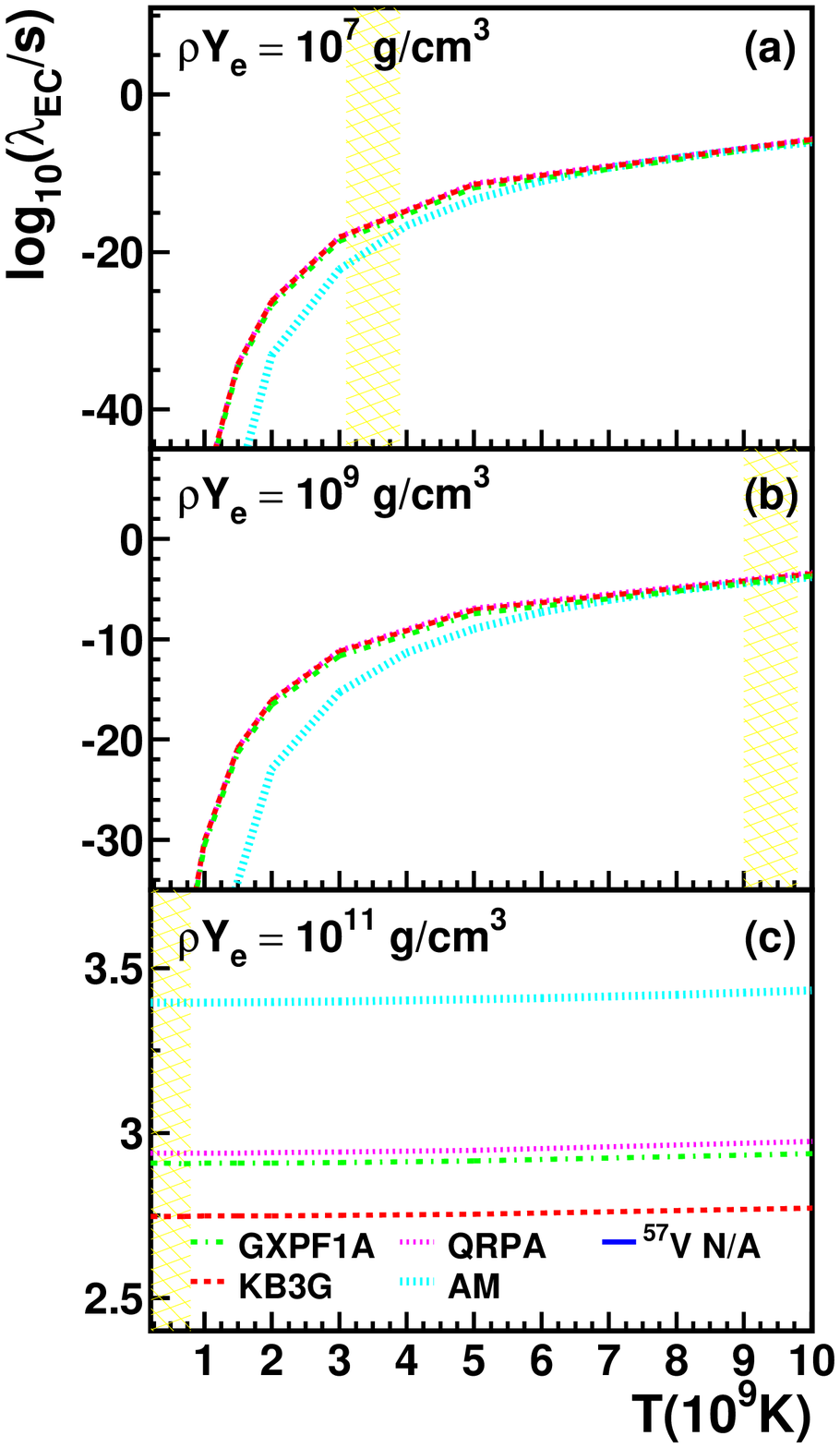}
 \includegraphics[bb = 0bp 0bp 385bp 667bp,clip,scale=0.315]{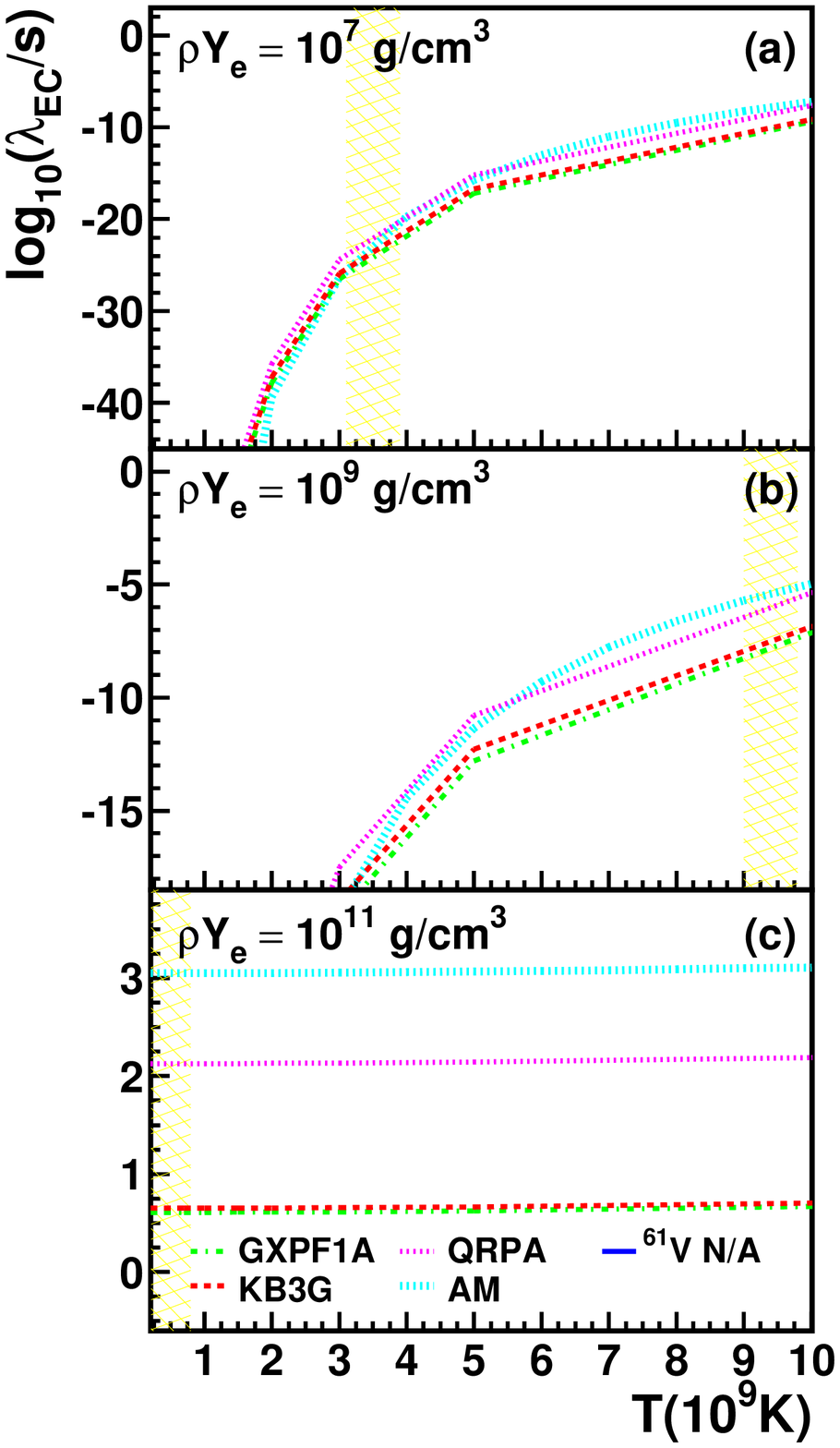}
 
 \caption{\label{fig4} The deduced B(GT) (top row) distributions (a-d) and sum strength (e) in select Ti nuclei, from ECs on the $^{43-62}$V isotopic chain, based on shell-model with (a) GXPF1a and (b) KB3G $pf$-shell effective interactions, (c) QRPA, and beta-decay or CE data (d), where available. Here we show example results from $^{54,57,61}$V$\rightarrow ^{54,57,61}$Ti cases (left, center and right, respectively). Beneath each panel (bottom row) are the corresponding EC rates, including that determined using the AM, as a function of temperature for stellar electron densities of (a) 10$^{7}$, (b) 10$^{9}$ and (c) 10$^{11}$ g/cm$^{3}$. The temperature regions used to determine select EC rate ratios for sensitivity analysis are highlighted in yellow cross-hatching and represent relevant temperatures for a variety of astrophysical scenarios.}
 \end{figure*}

 \begin{figure*}
 
\includegraphics[scale=0.31]{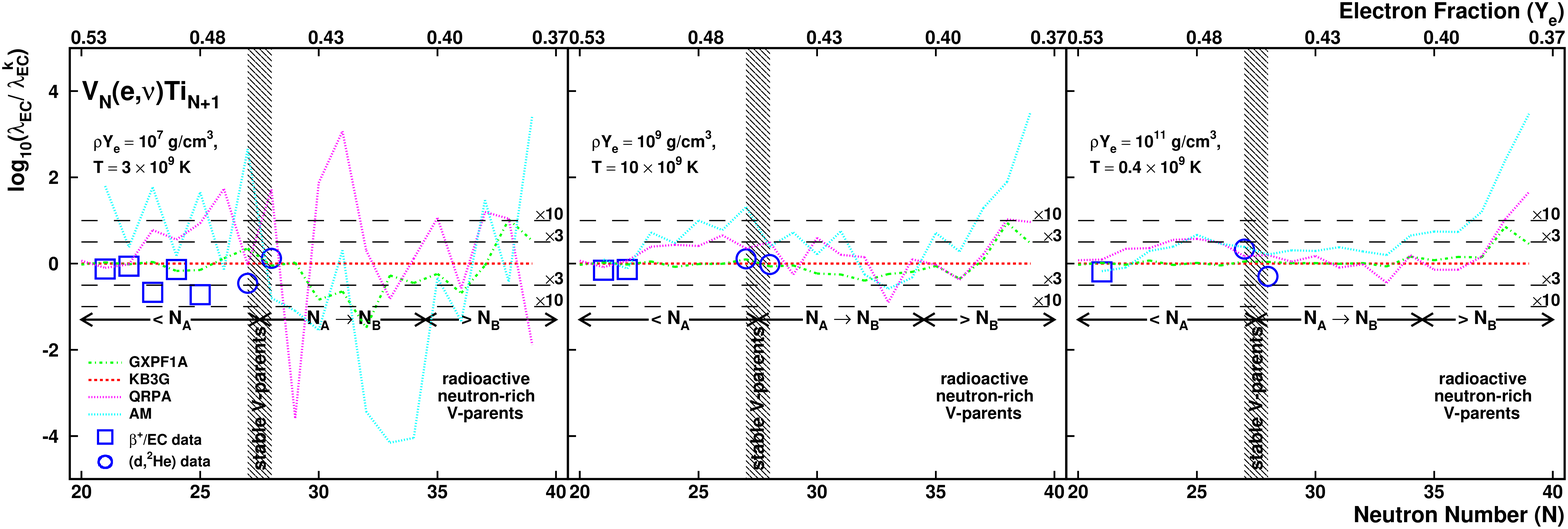}
 \caption{\label{fig5} The logarithm of the ratio (log-ratio) of EC rates versus KB3G-based rates, for GXPF1a rates (green, dot-dashed), QRPA rates (purple, dotted), AM rates (light-blue, dotted) decay data (blue squares) and CE data (blue circles), plotted versus parent neutron number and corresponding electron fraction for 10$^{7}$ g/cm$^{3}$ and 3$\times$10$^{9}$K (left), 10$^{9}$ g/cm$^{3}$ and 10$\times$10$^{9}$K (center) and 10$^{11}$ g/cm$^{3}$ and 0.4$\times$10$^{9}$K (right).}
 \end{figure*}

We now examine the rate sensitivities extracted from the three neutron number categories for statistically significant differences between them and make some observations on results from the representative set of isotopes. First, at the lowest density considered (Fig.\ref{fig3}, lower, all (a) panels), shell-model based rates present the smallest sensitivity because of their overall success reproducing the low-lying B(GT) distribution, relative to available data. As expected, QRPA determination of the low-lying B(GT) is relatively less consistent, and so, despite agreeing with EC rates based on data very well in some specific cases, has larger differences at the lower density and temperatures. By assigning a single, effective GT state in the daughter, the AM EC rates are the most sensitive at the lower densities and temperatures.

Consider $^{44}$V (Fig.\ref{fig3}, left), which is an example of a case where decay data is relatively useful since it is a large-$Q$, $Q_{>}$ case. When the TC is satisfied for a given thermodynamic condition, the effect of missing data above $E_{x} = Q$ in the daughter nucleus is negligibly small. Consequently, different B(GT) distributions (top panels, a-d) have nearly the same $T$-dependence for their associated EC rates (bottom panels, a-c) so long as each recovers the summed strength up to $E_{x}$. For example, in the middle rate panel (lower, b) for this isotope, rates based on data have different $T$-dependence because under these conditions the phase space is beginning to sample $E_{x} > 6$ MeV where the theoretical summed strengths (top, e) are higher than data.
 
For densities and temperatures such that a decay data set fails the TC, these are rejected from further analysis. It is noteworthy that what is meant by the `Q-value effect', when mentioned in the literature in connection with decay data, is where strength accessible due to thermodynamic condition is at $E_{x} > Q$, so that decay data and theoretical B(GT)s cease to be comparable. EC rate sensitivities extracted in such cases introduce systematic error. This is \textit{not} what is meant by sensitivity effects caused by the phase space changes from $Q_{>}$ to $Q_{<}$, as discussed earlier in this work in reference to Fig.\ref{qsample}. Here, the large phase space volume and more shallow exponential slope with respect to daughter $E_{x}$ in Fig.\ref{qsample} for a decay data case ($Q_{>}$) at $E_{x} < Q$ means the EC rate is dominantly determined by the summed strength and to a lesser degree by the lowest lying state. Uncertainty in the energy of low-lying states does not as significantly translate into rate uncertainty, as it does for $Q_{=}$/$Q_{<}$ cases. $^{44}$V is an example of this and at the lowest density, where sensitivity should be highest, the differences in rates between SM and QRPA cases is rather small because up to around 2 MeV, their summed strengths agree and the exact distribution below that is not as important.
 
The $^{50}$V case is unusual, since in terms of $Q$-value, it is a radioactive, proton-rich case, but because the decay to low-lying states in $^{50}$Ti is highly forbidden (6$^{+}$ entrance channel relative to the 0$^{+}$ $^{50}$Ti g.s.), $^{50}$V is stable and low-lying B(GT) that would otherwise ease the EC rate sensitivity does not exist in the daughter spectrum. This previous point highlights the importance of CE measurements for which a TC does not apply and where all GT strength that could participate in EC is occulted at higher excitation energy in $^{50}$Ti but is accessible from CE measurements. Consequently, for the AM based rate, which places all the B(GT) in a single, effective GT state, the $^{50}$V data gives a large log-ratio value and is responsible for a large share of the sensitivity factor and variance in the $<N_{A}$ group. This is the clearest for AM rates at $^{50}$V in Fig.\ref{fig5} in the lower two density conditions (left side of the cross-hatched, stable band). This is the isotope with the highest $N$ that still belongs to the $<N_{A}$ group.
 
The $^{51}$V case is an example typical of most CE data, having a stable parent and a small, but negative $Q$-value. This is also a case with the smallest parent $N$ and belonging to the $N_{A}\rightarrow N_{B}$ group. Previous work like Ref.\cite{PhysRevC.86.015809} has quantified the EC rate sensitivity for such cases ($N_{A}\rightarrow N_{B}$).
 
The $^{54,57,61}\textrm{V}\rightarrow ^{54,57,61}\textrm{Ti}$ cases, shown in Fig.\ref{fig4}, respectively, are cases with radioactive V parents for which EC decay is energetically forbidden and no CE measurements in inverse kinematics have ever been performed. Here then, the rates and sensitivities are pure predictions, with sensitivity factors with respect to data for $N_{A}\rightarrow N_{B}$ and $>N_{B}$ shown in Fig.\ref{fig8} being determined using properties of logarithms (e.g. $\log(\lambda_{\textrm{GXPF1a}}/\lambda_{\textrm{data}}) = \log(\lambda_{\textrm{GXPF1a}}/\lambda_{\textrm{KB3G}}) - \log(\lambda_{\textrm{data}}/\lambda_{\textrm{KB3G}})$) and propagating errors from the $<N_{A}$ values versus KB3G-based rates. $^{54}$V is also a case in the $N_{A}\rightarrow N_{B}$ group, but because the model space between various theories are fairly consistent with each other, most tend to converge on similar values for the B(GT) summed strength at sufficiently high $E_{x}$. The $^{54}\textrm{V}$ case is further noteworthy because, like other cases where the $Q$-value is small and negative, the EC rate is very sensitive to the input B(GT) and so the differences between theoretical rates, especially at lower density as shown in Fig.\ref{fig5} (left), are typically the largest. Specifically, as the temperature increases at lower densities of 10$^{7}$ g/cm$^{3}$ to 10$^{9}$ g/cm$^{3}$, as shown in Fig.\ref{fig4} (bottom, middle (a) and (b)), absolute rates increase by many orders of magnitude as the TC is suddenly fulfilled when $U_{F}$ first encounters GT states around 2-3 MeV in $^{54}$Ti (Fig.\ref{fig4} top, middle). This high sensitivity to the low-lying B(GT) results in 2-4 orders of magnitude differences between theories considered. Contrast this with the $^{57,61}$V cases, examples of $>N_{B}$ cases where different theories have significantly different degrees of freedom. Consequently, the summed strength does not tend to settle to similar values at high $E_{x}$. 

In the $^{61}$V case, it is clear that the expected additional degrees of freedom for QRPA lead to $\times100$ as much total strength in the daughter. In a simple IPM picture of the $pf$-shell valence, the neutron $1f_{5/2}$ single particle orbital is full and so GT transitions would be completely Pauli blocked. Thus, B(GT) in shell-model is solely due to residual interactions. In reality, close to an island of inversion, contributions from intruder orbitals are very likely, but of course QRPA is not limited by this. In these cases, our procedure for estimating sensitivities, like those shown for the $^{61}$V in Fig.\ref{fig5}, are likely lower boundaries for the actual value.

\begin{figure*}
\includegraphics[bb = 0bp 0bp 720bp 410bp,clip,scale=0.6]{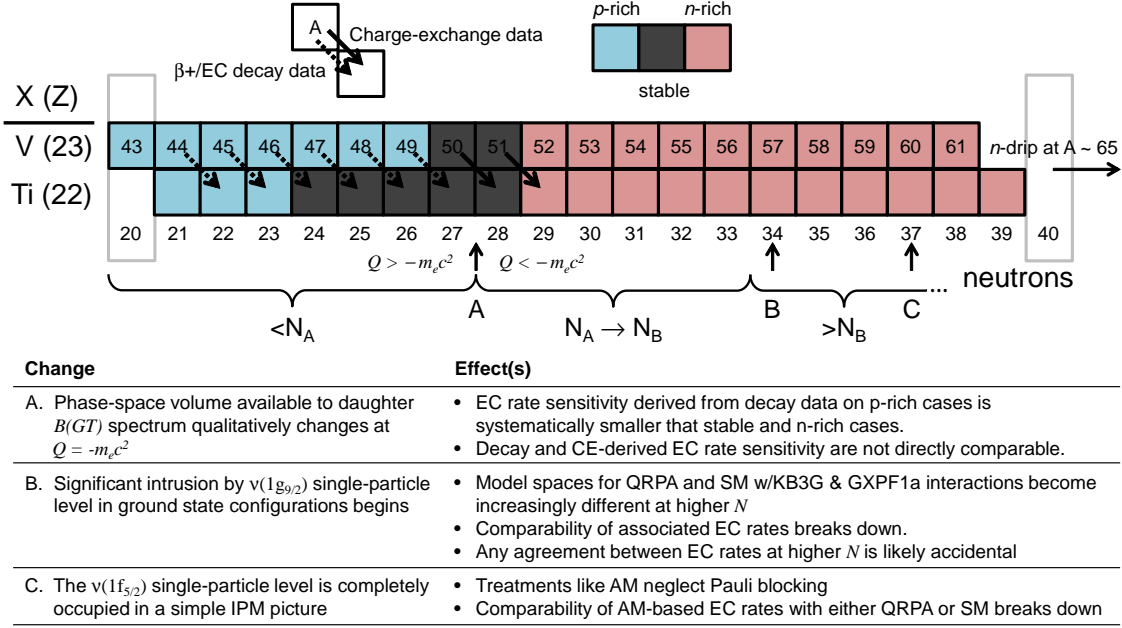}
\caption{\label{fig6} A chart of nuclides summarizing the cases considered for investigating influences on EC rate sensitvity to nuclear structure input.}
\end{figure*}

\begin{figure}
\includegraphics[bb = 10bp 30bp 430bp 713bp,clip,scale=0.5]{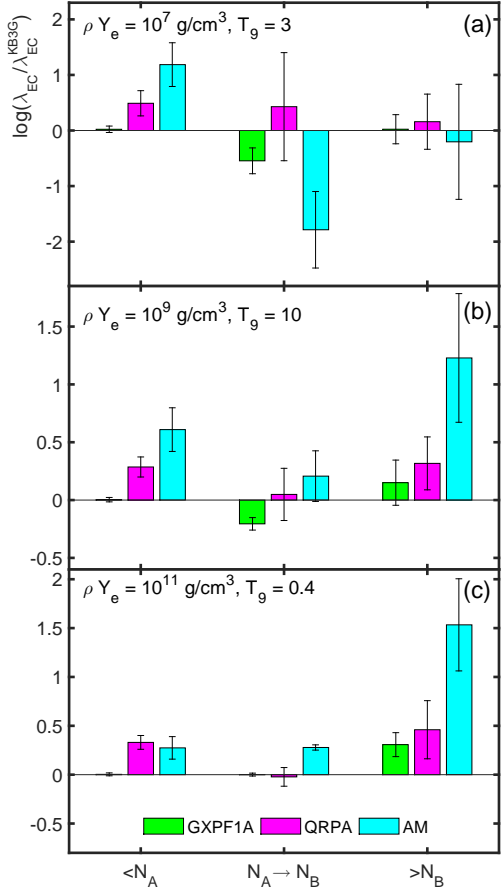}
\caption{\label{fig7} Average values of the log-ratio of EC rates versus KB3G-based rates, from left to right, averaged over the $<N_{A}$, $N_{A}\rightarrow N_{B}$ and $>N_{B}$ categories, respectively and at the representative thermodynamics conditions of 10$^{7}$ g/cm$^{3}$ and 3$\times$10$^{9}$K (a), 10$^{9}$ g/cm$^{3}$ and 10$\times$10$^{9}$K (b) and 10$^{11}$ g/cm$^{3}$ and 0.4$\times$10$^{9}$K (c). Error bars are standard errors in the mean ($\sigma/\sqrt{n}$).}
\end{figure}

\begin{figure}
\includegraphics[bb = 10bp 30bp 430bp 713bp,clip,scale=0.5]{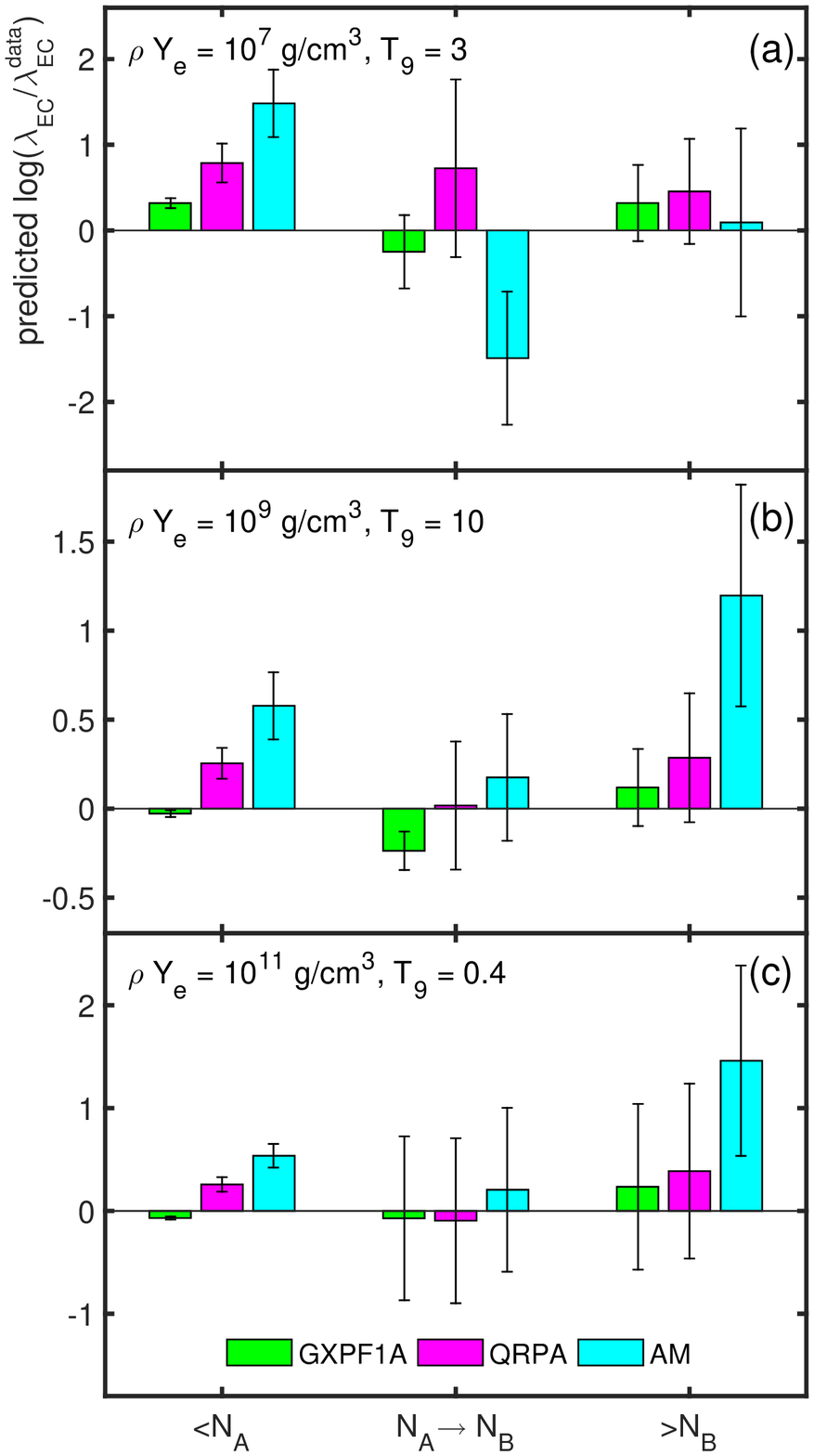}
\caption{\label{fig8} Average, predicted values of the log-ratio of EC rates versus data-based rates, from left to right, averaged over the $<N_{A}$, $N_{A}\rightarrow N_{B}$ and $>N_{B}$ categories, respectively and at the representative thermodynamics conditions of 10$^{7}$ g/cm$^{3}$ and 3$\times$10$^{9}$K (a), 10$^{9}$ g/cm$^{3}$ and 10$\times$10$^{9}$K (b) and 10$^{11}$ g/cm$^{3}$ and 0.4$\times$10$^{9}$K (c). Error bars are standard errors in the mean ($\sigma/\sqrt{n}$).}
\end{figure}

\begin{figure}
\includegraphics[bb = 10bp 0bp 430bp 237bp,clip,scale=0.5]{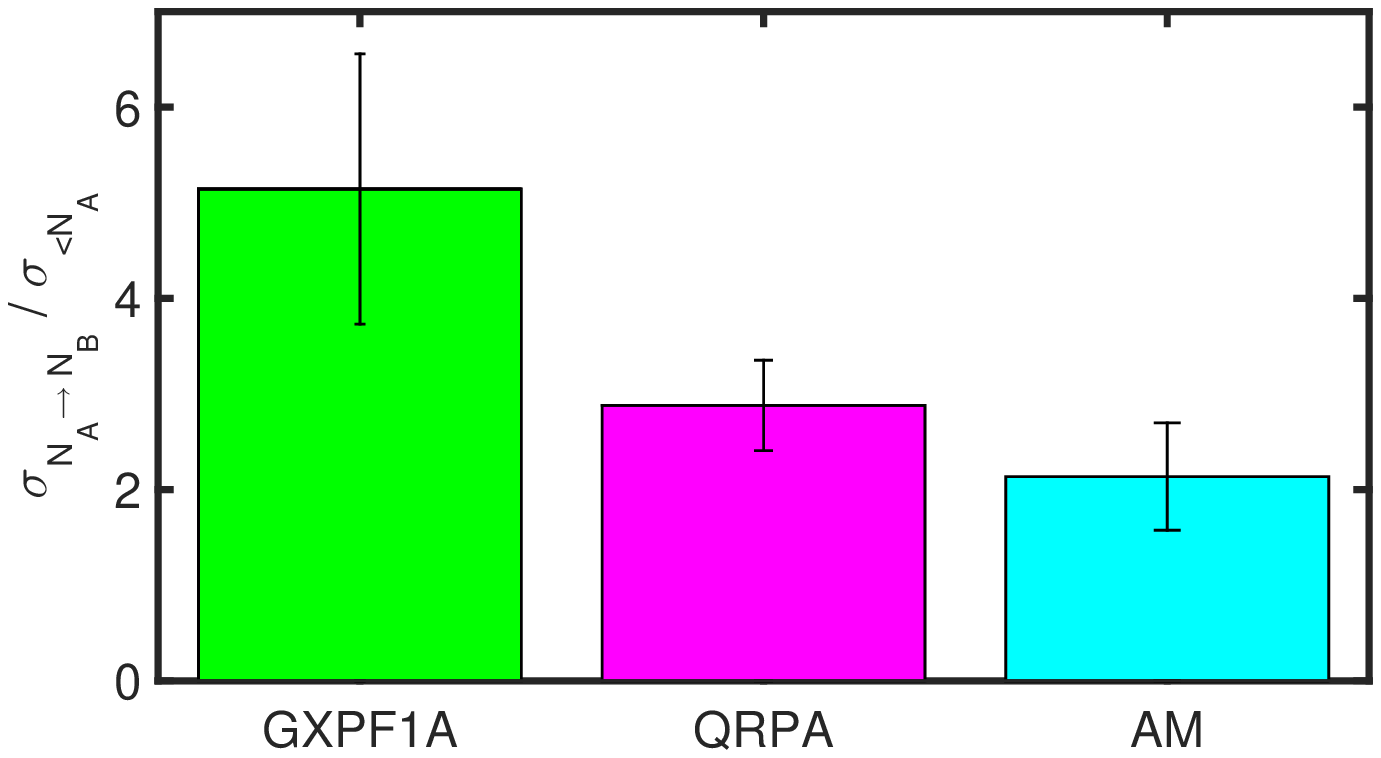}
\caption{\label{fig9} The factor by which standard deviations in log-ratios of EC rates increase, in the $N_{A}\rightarrow N_{B}$ versus the $<N_{A}$ groups.}
\end{figure}

Fig.\ref{fig6} shows a chart of nuclides for the V-Ti isotopic chain, color-coded for $p$-rich (cyan), $n$-rich (pink) and stable (grey) cases treated and labeled with arrows to show decay (dotted arrow) and CE (solid arrow) data used. Fig.\ref{fig6} also provides a schematic overview of the main results which we enumerate below. These are:

\begin{enumerate}

\item \textbf{Validation of the categories.} The systematic over-prediction, under-prediction, over-prediction pattern seen consistently and significantly in Figs.\ref{fig5}, \ref{fig7} and \ref{fig8} across increasing neutron-richness clearly show a need for separate consideration of the $<N_{A}$, $N_{A}\rightarrow N_{B}$ and $>N_{B}$ groups.

\item \textbf{Limited high-density convergence.} The expectation from the literature is that differences between EC rates based on data and based on theory should decrease with increasing density, so that the log-ratio of EC rates should tend toward zero. Looking left-to-right over Fig.\ref{fig5} panels or top-to-bottom along a category in Figs.\ref{fig7} and \ref{fig8} one can see convergence with increasing density. For the $<N_{A}$ and $N_{A}\rightarrow N_{B}$ groups, this is largely true. However, at high neutron-richness, represented by the $>N_{B}$ group, SM and QRPA log-ratios do not significantly change with increasing density, as seen in Fig.\ref{fig8}(a-c) down the right column of bars. At the highest densities, this is because the rate-ratio is set by the ratio of strengths and details of the B(GT) distribution no longer matter. For the AM log-ratios, the values actually increase with increasing density because its parameters are effective values based on fits to SM over many isotopes.

\item \textbf{Differences between decay and CE data.} Looking left-to-right within Fig.\ref{fig5} panels or left-to-right within Figs.\ref{fig7} and \ref{fig8} panels, once can see the behavior of log-ratios of EC rates with respect to increasing neutron-richness, with $<N_{A}$ (proton-rich), $N_{A}\rightarrow N_{B}$ (stable and moderately neutron-rich) and $>N_{B}$ (very neutron-rich) groups. Clearly, values of the log-ratio of EC rates extracted from decay data which is only available in the $<N_{A}$ group, are not predicted to hold at higher neutron-richness. Furthermore, the sensitivity factors derived from decay data are significantly different from those typically derived from CE data which is mostly applicable to $N_{A}\rightarrow N_{B}$ cases. This is a major result of the present report; that EC rate sensitivities determined from decay data are not immediately useful to estimate the same sensitivity at high neutron-richness without additional considerations. Even SM rates, going across categories, show significant differences at any of the three thermodynamic grid points investigated here. The same is true of QRPA and AM-based rates, as seen in Fig.\ref{fig8}.

\item \textbf{Growth of variances at high neutron-richness.} The statistical significance of differences between categories is complicated by an important fact: the variance/standard deviation among log-ratio of EC rates within each category is much larger for the neutron-rich categories than for the $<N_{A}$ category. Thus, not only do the sensitivity factors extracted from decay data, available only in the $<N_{A}$ category, significantly change at higher neutron-richness, their values are also less certain. For $<N_{A}$ cases, SM clearly out-performs other structure calculation methods, but with increasing neutron-richness, loses some of its advantage. Under some thermodynamic conditions examined here, the performance of SM and QRPA become more comparable, such as the high density, high neutron-richness cases, as in Fig.\ref{fig8}(b,c). Fig.\ref{fig9} shows the degree to which standard deviations of log-ratios increase from the $<N_{A}$ group to the $N_{A}\rightarrow N_{B}$ group. The corresponding change is negligible from the $N_{A}\rightarrow N_{B}$ to $>N_{B}$ categories. This change is driven by the changes to the phase space function with respect to daughter excitation energy, as discussed previously in Fig.\ref{qsample}.

\end{enumerate}

\section{Conclusion\label{conc}}

We have for the first time studied the stellar EC rates on the ground states of twenty V isotopes with 0.37 $\leq Z/A \leq$ 0.53. We have also for the first time considered the utility of B(GT) available from decay data. For the pre-collapse conditions of core-collapse supernovae (10$^{7}$ g/cm$^{3}$,T$_{9}=3$), we find sensitivity factors reaching 0.5-2 orders of magnitude, from shell-model-based to QRPA and AM-based rates, respectively, which are particularly disconcerting at high neutronization ($N\sim 40$). The consequences of such large uncertainties at the highest neutronizations, where maximum competition between weak rates and neutron reaction rates may occur, show the pressing need for CE reactions along strategic isotopic chains to low Z/N ratios at future RIB facilities. As recently shown in Ref.\cite{0004-637X-816-1-44}, factors smaller than these already have dramatic effects on inner core mass and neutrino luminosity. Toward the end of this isotopic chain at N$\sim$40, we expect that the sensitivities for any theoretical rate likely falls in a range as high as $\sim$1-3 orders of magnitude, as seen elsewhere along the chain where $Q < -m_{e}c^{2}$. This is not surprising, since there are essentially no experimental inputs to guide theory in the treatment of shape deformation, shape-coexistence and intruder orbits at such high $N$. The situation improves significantly as one progresses to higher densities and the EC rate sensitivity to details of the low-lying B(GT) substantially abate as shown in Fig.\ref{fig5} (right). Only at such higher densities do EC rate sensitivity factors of 0.5-2 orders of magnitude, which are often taken in the literature (e.g. Ref.\cite{10.1038.nature.12757} for neutron star crustal processes), appear valid.

Finally, we further emphasize the need for additional CE measurements at future RIB facilities, motivated by two new, important results. First, that sensitivities extracted from decay data on $p$-rich parents significantly underestimate what should be expected for the $n$-rich side because of the different energy dependence of the phase space. Second, theoretical comparisons to CE measurements in forward kinematics on stable targets only provide sound estimates of the EC rate sensitivity up to a certain boundary value $N = N_{B}$ unique to each chain. This defines a zone from stability to several neutrons away from stability and from $Z = 20$ to $Z = 40$ (applying the definition of $N_{A}\rightarrow N_{B}$ at successive $Z$-values) where results like those in Ref.\cite{PhysRevC.86.015809} are confirmed. But beyond this, theories diverge on account of model space inconsistencies. B(GT) is currently not available for $N > N_{B}$ cases to guide improvements and there is now a need for creating a method for improvements to theory to be applied directly or extrapolated from cases at lower neutron-richness. With $\sim$100 neutron-rich isotopes in the $pf$-shell, under the same astrophysical conditions one can expect $\sim$25 of them, based on the definition of the $>N_{B}$ group and so biased toward those at high neutron-richness, to have larger differences with theory than those found here. New CE measurements in inverse kinematics and new B(GT) calculations in very large model spaces are crucial for providing insight on $N > N_{B}$ cases throughout the $pf$-shell.

\begin{acknowledgments}
SG would like to acknowledge enriching discussions with Peter M\"oller. This work is supported in part by NSF grant PHY-1404442.
\end{acknowledgments}

\bibliography{reference}

\end{document}